\documentclass[epj]{svjour}
\usepackage[T1]{fontenc}
\usepackage{t1enc}

\usepackage{graphics}
\usepackage[dvips]{graphicx}
\usepackage{graphicx}
\usepackage{amssymb}
\usepackage{amsmath}
\usepackage{amsfonts}
\usepackage{amsbsy}
\usepackage{amscd}
\usepackage{amsopn}
\usepackage{amstext}
\usepackage{amsxtra}
\usepackage{color}
\usepackage{epsfig}
\usepackage{slashed}

\newcommand{\be}{\begin{equation}}
\newcommand{\ee}[1]{\label{#1} \end{equation}}
\newcommand{\ba}{\begin{eqnarray}}
\newcommand{\ea}[1]{\label{#1} \end{eqnarray}}
\newcommand{\nl}{\nonumber \\}

\newcommand{\REF}[1]{Eq.~(\ref{#1})}

\begin{document}
\title{Scale dependence of the $q$ and $T$ parameters of the Tsallis distribution in the process of jet fragmentation}

\author{Karoly Urmossy\inst{1} \thanks{\emph{email address:} karoly.urmossy@gmail.com}\and Antal Jakovac\inst{2}% etc
}                     % Do not remove
\institute{Department for Atomic Physics, Eotvos Lorand University,\\
1A Pazmany Peter Walk, H-1117 Budapest, Hungary \and Department of Computational Sciences, Wigner Research Centre for Physics,\\
29-33 Konkoly-Thege Mikloos Street, H-1121 Budapest, Hungary}
\date{Received: date / Revised version: date}
\titlerunning{Scale evolution of Tsallis parameters}
\authorrunning{K. Urmossy \and A. Jakovac}

\abstract{
The dependence of the $q$ and $T$ parameters of the Tsallis-distribution-shaped fragmentation function (FF) on the fragmentation scale (found to be equal to the jet mass) is calculated via the resummation of the branching process of jet fragmentation in the leading-log appriximation (LLA) in the $\phi^3$ theory. Jet and hadron spectra in electron-positron ($e^+e^-$) annihilations with 2- and 3-jet final states are calculated using virtual leading partons. It is found that jets, produced earlier in the branching process, are more energetic, and the energy, angle and multiplicity distributions of hadrons stemming from them are broader. It is also found that replacing the LL resummation in the branching process by a single splitting provides good approximation for the jet energy distribution in 2-jet events. Furthermore, a micro-canonical statistical event generator is presented for the event-by-event calculation of hadron momenta in $e^+e^-$ annihilations.
 % end of PACS codes
} %end of abstract
\maketitle
\section{Introduction}
\label{sec:intro}
The spectra of hadrons produced in high-energy collisions are more-or-less well described by various versions of phenomonological models based on the \textit{cut-power law} or \textit{Tsallis-distribution} (TS), $f_{TS}(E) \propto [1+(q-1)E/T]^{-1/(q-1)}$. $E$ is the energy of the hadron, and $q$ measures the deviation from the Boltzmann-Gibbs (BG) distribution ($q=1$). The dependence of the $q$ and $T$ parameters on the features of the collisions ($\sqrt s$, centrality, type and number of the produced final state particles) has extensively been studied recently in electron-positron ($e^+e^-$) \cite{bib:Becattini}--\cite{bib:UKee}, positron-proton ($e^+p$) \cite{bib:FHLiu10}--\cite{bib:UKdis}, proton-proton ($pp$) \cite{bib:Trainor}--\cite{bib:Zhangbu2}, proton-nucleus ($pA$) and deuteron-nucleus (dA) \cite{bib:FHLiu2}--\cite{bib:Zhangbu2}, nucleus-nucleus (AA) \cite{bib:FHLiu2}--\cite{bib:WCZhang7} collisions, and Drell-Yan processes \cite{bib:FHLiu14}. The found tendencies as well as the quality of the fits to measured data depend on the version of the model used. Some models conjecture that quark-gluon plasma (QGP), an expanding and cooling thermal source is created in the examined collision, and use the Cooper-Frye formula to calculate the distribution of hadrons produced at break-up \cite{bib:FHLiu5}--\cite{bib:Zhangbu3}, \cite{bib:LLiu}--\cite{bib:Zhangbu2}, \cite{bib:FHLiu7}--\cite{bib:UKaa}, \cite{bib:FHLiu15}--\cite{bib:Zhangbu4}. These models provide a better agreement with measured data in $AA$ collisions, especially in the interval, where the hadron's transverse momentum is smaller than its mass $p_T\lessapprox m_h$. However, in order to describe both the spectra and the azymuthal anisotropy $v_2$ in AA collisions at $\sqrt s\geq 200$ GeV, and $p_T$ up to 20 GeV/c, two-component models \cite{bib:FHLiu13}--\cite{bib:Zhangbu4} are needed. Such models make out hadron yields from those hadrons, which stem from the QGP (\textit{'soft'} component), and those, which stem from jets (\textit{'hard'} component). It has been found that the more central the collisions (the greater the hadron multiplicity), the larger the ratio of the \textit{soft} hadrons and the closer their TS to the BG distribution. This suggests that in more central collisions, the QGP is closer to equilibrium \cite{bib:WCZhang7}--\cite{bib:Zhangbu4}. The distribution of the \textit{hard} yields, which dominates the spectrum and $v_2$ for $p_T\gtrapprox$ 5 GeV/c, agrees with spectra measured in peripheral AA, or pp collisions. Its $q$ parameter increases with $\sqrt s$ and hadron multiplicity. Quark-coalescence models \cite{bib:UKaa} suggest that the power of hadron spectra scales with the number of constituent quarks, thus, the $q$ parameters of mesons and baryons obey the relation $(q_M-1)/(q_B-1) = 3/2$. Although, this precise ratio has not been observed, it was found that $q_M > q_B$. Besides, the $T$ parameter was found to depend linearly on the hadron mass. In case of $e^+e^-$ annihilations, hadron spectra are measured up to a range, where hadron energies are comparable with $\sqrt s$, thus, micro-canoncal statistical models are used \cite{bib:Becattini}--\cite{bib:Begun3} for the description of fragmentation functions ($FF$) and event shapes. The canonical statistical version of the TS distribution fails to describe the spectra of jets \cite{bib:LNGao,bib:FHLiu12} and direct photons \cite{bib:FHLiu1} for $p_T\gtrapprox0.1\sqrt s$ in $pp$, $pA$ and $AA$ collsions. 

From the theoretical point of view, the TS distribution has been obtained in the canonical and micro-canoncal ensembles in case when the temperature \cite{bib:Beck,bib:Wilk1}, volume \cite{bib:Begun2,bib:Begun3} or multiplicity \cite{bib:BiroNflukt} fluctuates according to specific patterns due to some external reasons. The TS is also the steady solution of the Langevin and Fokker-Planck equations with damping and noise terms depending linearly on the particle's energy \cite{bib:BiroJako}. Besides, thermodynamic properties of the \textit{Linear Sigma Model} \cite{bib:Castano}, the \textit{Nambu-Joana-Lasinio Model} \cite{bib:YPZhao}, a hadron gas in the presence of magnetic field \cite{bib:Pradhan} along with general thermodynamic relations \cite{bib:Ishihara} and a small $q-1$ expansion have been obtained in the framework of the TS statistics. It is, however, important to point out that a \textbf{\textit{solid derivation of the TS distribution from first principles of the strong interaction is still missing}}. Nevertheless, it has been shown in \cite{bib:CYWong} that the $p_T$ spectrum of a leading parton (jet) obtained from the leading order QCD cross section of two-parton scatterings in $pp$ collisions, approximately takes the form $d\sigma^{\,Jet}/dx \approx (1-x)^{a}/x^{4.5}$, with $x = 2p_T/\sqrt s$. The $(1-x)^a$ factor comes from the form of the most commonly used parton distribution functions (PDF), and the $x^{-(4+1/2)}$ factor stems from the hard scattering of partons in the incoming protons and the integration for the not measured other jet momenta. This result has estimated the power of jet spectra in accordance with Tevatron and LHC results, and it has also accounted for the rapid decrease of the distribution for large $p_T$, where the jet energy becomes comparable with $\sqrt s$. In the paper, it is also argued that the low-$p_T$ behaviour of jet spetra, where multiple scattering and non-perturbative processes become important, cannot be explained based on the single hard scattering of partons. These arguments suggest that the $q$ parameter of the TS may be calculated in perturbation theory (PT), but the $T$ parameter is of non-perturbative origin.

To avoid having to deal with the parton structure of the colliding particles, in this paper, we examine $e^+e^-$ annihilations, and derive the scale dependence of the $q$ and $T$ parameters of the TS shaped hadron spectrum in fragmentation processes. Inspired by \cite{bib:Graz,bib:p3dglp,bib:CollinsMC}, we use the $\phi^3$ model, which is the simplest asymptotically free quantum field theory (QFT) mimicing the 3-gluon vertex of QCD. As perturbative methods for the calculation of the fragmentation of an off-shell leading parton to hadrons need a non-perturbative input, the form of the fragmentation function (FF) at a low starting scale, which cannot yet be derived from first principles, there is room for model building at this point. For this purpose, we use the model \cite{bib:UKpp3D,bib:UKdis} presented in Sec.~\ref{sec:D0}, because it takes into account the finitness of the total energy of the produced hadrons using micro-canonical statistics, and it obtains the TS distribution making use of the experimentally observed negative binomial (NBD) hadron multiplicity fluctuations. Similarly to \cite{bib:p3dglp,bib:QCDdglp}, in Sec.~\ref{sec:evol}, we start from the Dyson-Schwinger equation (DSE) for the fragmentation of a single parton to hadrons, we put momenta of daughter partons onto the mass-shell, and sum up the leading-log (LL) terms via the Dokshitzer-Gribov-Lipatov-Altarelli-Parisi (DGLAP) equation \cite{bib:DGLAP}. We obtain the dependence of the $q$ and $T$ parameters on the virtuality of the leading parton (scale) via solving the DGLAP equation for the moments of the FF. In Sec.~\ref{sec:res}, we calculate jet and hadron distributions in $e^+e^-$ annihilations with 2-, and 3-jet final states using a Monte-Carlo (MC) event generator method based on the previously obtained FF. We summarize the results in Sec.~\ref{sec:fact}.

The approach presented in this paper mainly differs from most pQCD methods in that it allows for the virtualities of the leading partons to be arbitrarily large (within the boundaries set by energy-momentum conservation), whereas usual parton model calculations use on-shell leading partons, as they rely on factorisation theorems (FT). Consequently, we obtain broad distributions for jet masses and for the angles between jet and hadron momenta, unlike in works like \cite{bib:dEnterria1} which use a kinematic approximation in which, momenta of leading and daughter partons are nearly collinear. FTs \cite{bib:fact,bib:fact2} prove that if jets are highly boosted bunches of particles of low total momentum squared ($M_J = \sqrt{P_J^2} \ll E_J$), and the angles between jets are large, then, their contribution to cross sections can be expressed by multiplicative factors (convoluted with the hard part of the process). Interference terms coming from gluon exchange between jets or jets and soft processes can either be incorporated into the jet factors via Wilson-lines, or can be neglected, as they are suppressed by factors of $M_J/E_J\ll1$. However, in a portion of events, the basic conditions of FTs do not hold, as jet masses are comparable to jet energies according to measurements \cite{bib:MjetEE1,bib:MjetEE2,bib:MjetPP}. 

Our approach resembles the recursive method presented in \cite{bib:CollinsMC} which generates virtual daughter partons in each round, however, in two steps. In the first step, it generates on-shell daughter partons using a fixed-order cross sections, then in the second step, it generates daughter parton 'masses' using the distribution obtained from the fixed-order cross section of on-shell grand-daughter production. This procedure starts again in the next round, until parton virtualities decrease to the order of hadron masses. This way, every parton in a given generation of the cascade may be virtual, however, the vector part of their momenta are generated according to cross sections involving mother and daughter partons, while virtualities are obtaied from cross sections involving daughter and grand-daughter ones. In our method on the other hand, only the leading parton of a jet is virtual, as it produces on-shell daughters in the fragmentation process, but both the vector part of its momentum and its 'mass' are generated from the same distribution obtained from the LL resummation of the fragmentation process (not from fixed-order graphs, as in \cite{bib:CollinsMC}). 

\onecolumn
\section{A statistical model for the FF at starting scale}
\label{sec:D0}
We use the minimalistic conjecture, that at the starting scale $M_0$, the sole constraint required throughout the hadronisation process is the conservation of energy-momentum. This way, hadrons stemming from the leading parton of momentum $P=\left(\sqrt{M_0^2 + \mathbf{P}^2},\mathbf{P}\right)$ form a micro-canonical ensemble. In the micro-canonical ensemble, the main quantity, which determines the distribution of particles, is the phasespace, which, for $n$ massless particles in $D$ dimensions is
\be
\Omega_n(P) \;=\; \prod_{i=1}^n \int \frac{d^{D-1}\mathbf{p}_i}{p_i^0}\, \delta^{D}\left(\sum_j p_j^\mu-P^\mu \right) \sim M_0^{n(D-2)-D}\;.
\ee{mic1}
This result follows from dimensional analysis and Lorentz-invariance, but the detailed calculation can be found in Appendix~\ref{sec:micro}. Consequently, the one-particle distribution in the micro-canonical ensemble is
\be
d_n(x,M_0) \;=\; \frac{\Omega_{n-1}(P-p)}{\Omega_n(P)} =  \frac{\left(\frac{2}{M_0}\right)^{\omega^\ast} \Gamma\left(\frac{\omega^\ast n}{2}\right)}{\kappa_{D-1}\Gamma(\omega^\ast˘) \Gamma\left[\frac{\omega^\ast (n-2)}{2}\right]}  \left(1 - x\right)^{\frac{(n-2)\omega^\ast}{2} -1 }\;,
\ee{mic2}
with $x = 2pP/P^2$, $\omega^\ast = D-2$ and $D>2$. For example, the properly normalized distributions in $D=4$ and 6 dimensions are
\be
d_n(x,M_0) \;=\; \left\lbrace {{ \frac{(n-1)(n-2)}{\pi M_0^2}(1 - x)^{n-3}\, ,\qquad\textrm{if}\; D=4} \atop{
 \frac{(2n-1)\dots(2n-4)}{\pi^2 M_0^4}(1 - x)^{2n-5}\,,\qquad\textrm{if}\; D=6 \;.}} \right.
\ee{mic3}
In $D=4$ dimensions, for $n>3$, the probability of a particle to acquire the maximal energy (which is $M_0/2$ in the co-moving frame) is zero, as $d_n(1,M_0)=0$. The $n=3$ case is exceptional, as $d_3(x,M_0)$ is a uniform distribution. In the $n=2$ case, $d_2(x,M_0)\sim\delta(1-x)$, and energy-momentum conservation cannot be satisfied for $n<2$.  Interestingly, the same variable $x$, which is used in parton model calculations, emerges as the natural variable of the micro-canonical ensemble.

As measured fragmentation functions are averaged over the fluctuation of hadron multiplicity, we make use of the experimental observation \cite{bib:NjetPP,bib:UKee} that the multiplicity of hadrons in a jet fluctuates according to the negative-binomial distribution (NBD) 
\be
\mathcal{P}_n \;=\; {{n+r-1}\choose{n}} \bar{p}^n (1-\bar p)^r\;,
\ee{mic4}
with parameters $\bar p,r$, expectation value $\bar n = r\bar p/(1-\bar p)$ and scattering $\sigma^2 = r\bar p/(1-\bar p)^2$. Thus, we use the one-particle micro-canonical distribution averaged over multiplicity fluctuations as fragmentation function at starting scale $M_0$:
\ba
d\left(x,M_0^2\right) &\;=\;& \sum_n \mathcal{P}_n\,n\,d_n(x,M_0) =\nl
&=& \frac{2 \left(\frac{2}{M_0}\right)^{\omega^\ast} (1-\bar p)^r}{\kappa_{D-1}\Gamma(\omega^\ast+1) } \left(-\frac{\partial}{\partial x}\right)^{\omega^\ast+1}   \sum_n  {{n+r-1}\choose{n}} \bar p^n (1 - x)^{n\omega^\ast/2} \nl
&=& \frac{2 \left(\frac{2}{M_0}\right)^{\omega^\ast} (1-\bar p)^r}{\kappa_{D-1}\Gamma(\omega^\ast+1) }    \left(-\frac{\partial}{\partial x}\right)^{\omega^\ast+1}  \left[1 - \bar p(1 - x)^{\omega^\ast/2}\right]^{-r} \;.
\ea{mic5}
The evaluation of this formula in $D=4$ dimensions results in the Tsallis distribution
\be
d(x,M_0^2) \;=\; \left(\frac{\bar p}{1-\bar p}\right)^3 \frac{r(r+1)(r+2)}{\pi M_0^2} \left(1 + \frac{\bar p}{1 - \bar p} x\right)^{-(r+3)} \,,
\ee{mic6}
in which, the parameters of the multiplicity distribution expressed with the usual $q$ and $\tau$ parameters are
\be
r +3 \;=\; \frac{1}{q-1}\;, \qquad \frac{\bar p}{1-\bar p} = \frac{q-1}{\tau }\;. 
\ee{mic7}
The usual tempereature is $T = \tau M_0/2$.

Note that, when calculating \REF{mic6}, the $n=2$ term $d_2(x,M_0)$ was left out of the sum in \REF{mic5}. Besides, the TS distribution does not go to zero at $x=1$, due to the uniform distribution $d_3(x,M_0)$ which was included in the sum in \REF{mic5}. If we leave out the $n=3$ term from the sum, we arrive at a modified version of the TS distribution
\be
d^\ast(x,M_0^2) \;=\; \left(\frac{\bar p}{1-\bar p}\right)^3 \frac{r(r+1)(r+2)}{\pi M_0^2} \left[\left(1 + \frac{\bar p}{1 - \bar p} x\right)^{-(r+3)} - \left(1 + \frac{\bar p}{1 - \bar p}\right)^{-(r+3)} \right] \,,
\ee{mic8}
which obeys $d^\ast(1,M_0^2)=0$, thus, it decreases more rapidly than the TS function for $x\geq0.1$ providing a better description of hadron spectra in $e^+e^-$ annihilations, and jet spectra in $pp$ collisions.

\section{Scale evolution of the FF}
\label{sec:evol}

\begin{figure}%[!h]
\begin{center}
\includegraphics[width=0.7\textheight]{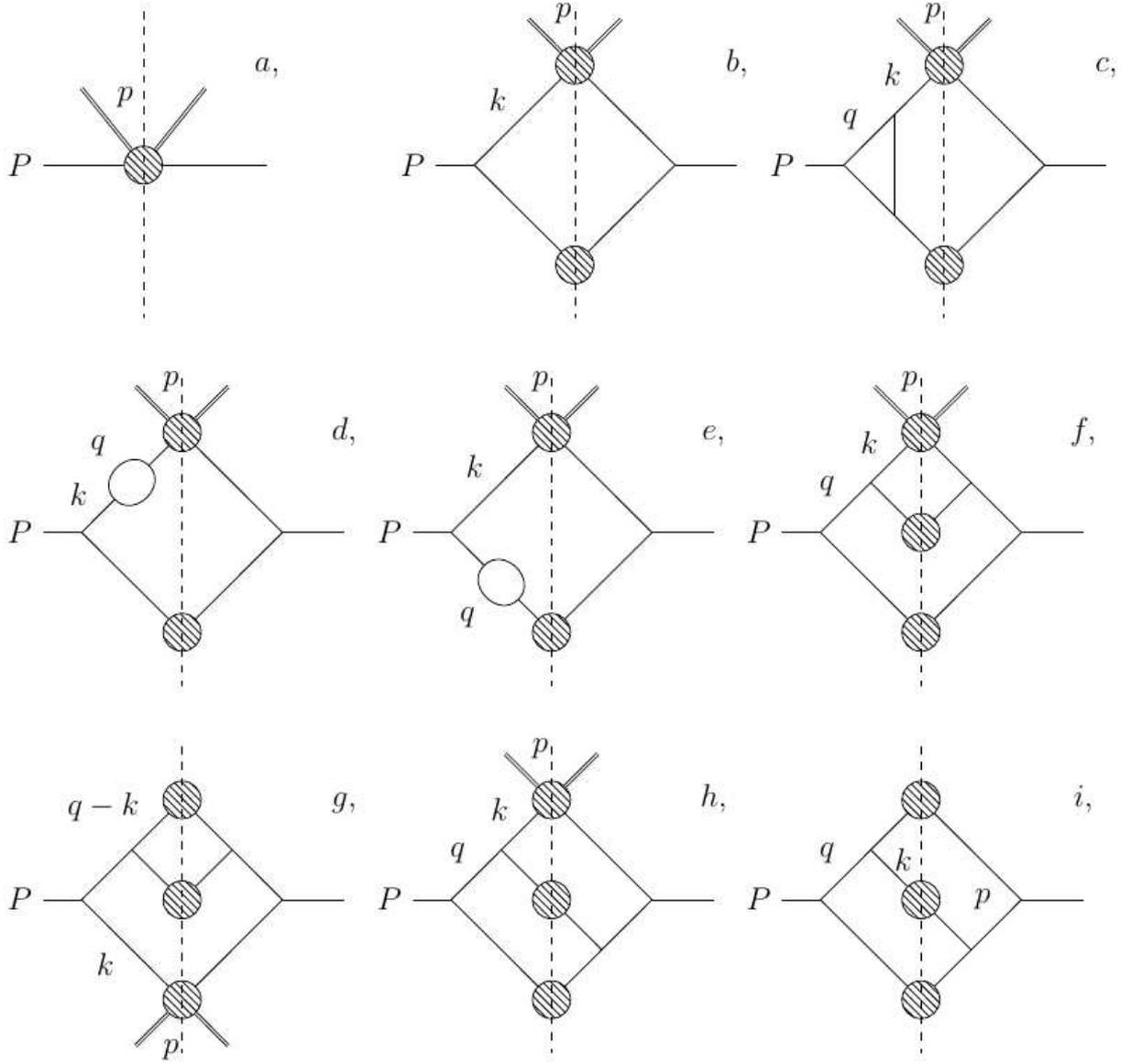} %PLB
\end{center}
\caption{Graphs contributing to the fragmentation of a virtual parton of momentum $P$ into a measured hadron of momentum $p$ plus anything at order $g^4$ in the $\phi^3$ theory.
\label{fig:DSE}}
\end{figure}

To obtain $D(p,P)$, the probability of a parton of momentum $P$ producing a hadron of momentum $p$ (the amputated cut graph blob in Fig.~\ref{fig:DSE}.a) at higher scales $P^2\geq M^2_0$, we use the DSE at $\mathcal O(g^4)$:
\ba
D(p,P) &\;=\;& g^2\int\frac{d^Dk}{(2\pi)^D}\, \frac{D(p,k) \bar D(P-k) }{k^4 (P-k)^4 } \left[1 +  g^2\int\frac{d^Dq}{(2\pi)^D} \left\lbrace \frac{i\, n_c}{q^2(q-k)^2(P-q)^2}\right.\right. +\nl
&&+ \left. \frac{i\, n_d}{k^2q^2(q-k)^2}  + \frac{i\, n_e}{(P-k)^2q^2(P-q-k)^2} \right\rbrace + \left. \bar n_c \left(\delta Z_g + \frac{3}{2}\delta Z_3\right) - (\bar n_d + \bar n_e)\delta Z_3 \right] \nl
&+& g^4\int\frac{d^Dk}{(2\pi)^D}\int\frac{d^Dq}{(2\pi)^D} \, \frac{D(p,k) \bar D(q-k) \bar D(P-q)}{k^4 (q-k)^4 (P-q)^4} \times\nl 
&&\qquad\qquad\qquad\times \left\lbrace \frac{n_f}{q^4} + \frac{n_g}{(P-k)^4} + \frac{n_h}{ q^2 (P-k)^2} + \frac{n_i}{q^2 (P-q+k)^2} \right\rbrace
   \;.
\ea{evol1}
Factors like $\bar D(k) = \int \frac{d^{D-1}\mathbf{p}'}{p'_0} D(p',k)$ arise when we integrate out the momenta of un-measured hadrons stemming from other jets. The combinatorial factors $n_c,\dots,n_i$ multiply terms depicted in Fig.~\ref{fig:DSE}.c--i.

If we were only interested in the partonic part of the branching process, $D(p,P)$ would denote the distribution of a \textit{parton} of momentum $p$ created by the leading parton of momentum $P$. In that case, the $p-$integrated cut blobs would be replaced by the cut propagator: $\bar D(k)/k^4\rightarrow\delta(k^2)$ at leading order, and we would arrive at the problem presented in \cite{bib:p3dglp}.

Let us parametrize the fragmentation function $D$ for later model building purposes as a product
\be
D(p,P) \;=\; P^{10-D} \rho\left(P^2\right) d\left(\frac{2pP}{P^2},P^2\right) 
\ee{evol2}
of a \textit{``virtuality (or jet mass) distribution''} $\rho\left(P^2\right)$ normalized as $\int dP^2\rho\left(P^2\right) = 1$, and a \textit{``hadron momentum distribution''} $d\left(x,P^2\right)$ normalized as $\int \frac{d^{D-1}p}{p_0} d\left(\frac{2pP}{P^2},P^2\right)=\bar n(P^2)$, being the average hadron multiplicity in the jet. The $P^{10-D}$ factor renders the mass-dimension $[D(p,P)] = 10-2D$ of the FF, and also removes the poles coming from the propagators on both sides of the cut blobs, as we work in $D=6$ dimensions.

Now, let us simplify the DSE via making the approximation of pushing the virtualities of the daughter partons (cut blobs) down to a negligible value $m_0^2\ll P^2$. It is interesting that the phasespace of a massless hadron of momentum $p$ stemming from a leading parton of momentum $k = (k_0,\mathbf k) = (k_0,k,\mathbf{k}_T)$ is an ellipsoid of center $\mathbf k/2$, length $k_0$ and width $m = \sqrt{k^2}$. This can be seen from the condition defining the boundary of the phasespace, which is the requirement that the phasespace of the rest of the hadrons stemming from the same jet $\Omega_{n-1}(k-p)$ needs to be positive. According to \REF{mic1}, it requires that $(k-p)^2\geq0$, or equivalently, $2pk/k^2 \leq1$. For instance, using the parametrisation $P=(M,0,\mathbf0)$, $k = (k_0,k,\mathbf 0)$ and $p = \left(\sqrt{p_\parallel^2+p_\perp^2},p_\parallel,\mathbf{p}_\perp\right)$, we arrive at the equation of an ellipsoid: $\left(\dfrac{p_\perp}{m/2}\right)^2 + \left(\dfrac{p_\parallel-k/2}{k_0/2}\right)^2\leq1$. If we push the virtuality of the hadronizing parton down to a negligible value $m\rightarrow m_0\approx 0$, the phasespace for the daughter hadrons shrinks to a one-dimensional interval as $p_T\rightarrow0$, $p\rightarrow(p_\parallel,p_\parallel,\mathbf0)$ with $p_\parallel\in[0,k_0]$. Besides, the first argument of $d(p,k)$ simplifies as $\dfrac{2pk}{k^2}\rightarrow \dfrac{2p_\parallel}{k_0+k}\rightarrow \dfrac{p_\parallel}{k}= \dfrac{x}{z}$, with $x=\dfrac{2pP}{P^2}=\dfrac{2p_\parallel}{M}$ and $z=\dfrac{2kP}{P^2}=\dfrac{2k}{M}$. We may carry out the above procedure in the SDE via setting
\be
\rho(\dots)\rightarrow (2\pi)\delta(\dots),\quad\textrm{and}\quad  d\left(\frac{2pk}{k^2},k^2\right) \rightarrow (2\pi)^{D-2} \delta^{D-2}(\mathbf{k}_T) \,  d_0\left(\frac{x}{z},m_0^2\right) \,,
\ee{evol3}
where $\mathbf{k}_T \mathbf{p} =0$. This way, %the integrals with respect to $k$ simplify as
%\ba
%\int \frac{d^Dk}{(2\pi)^D}\, (p,k)\rho\left(k^2\right) \dots &\rightarrow& \frac{1}{2(2\pi)} \int \frac{dz}{z}\, d_0\left(\frac{x}{z},m_0^2\right) \dots\;,\nl
%\int \frac{d^Dk}{(2\pi)^D}\, d(p,k)\rho\left(k^2\right) \rho\left[(P-k)^2\right] \dots &\rightarrow& \frac{1}{2} \frac{1}{M^2}\int \frac{dz}{z}\, d_0\left(\frac{x}{z},m_0^2\right) \delta(1-z) \dots\;,
%\ea{1p5}
and SDE becomes 
\ba
 P^2 D\left(x,P^2\right) &\;=\;& \frac{g^2}{2}\, d_0(x,m_0^2) \;+\; \frac{g^4}{2}\int\frac{dz}{z}\, d_0\left(\frac{x}{z},m_0^2\right) A(z,P^2)\;,\;\;\textrm{with}\nl
A(z,P^2) &\;=\;& \delta(1-z) \left[\int\frac{d^Dq}{(2\pi)^D} \left\lbrace \frac{i\,n_c}{q^2(q-k)^2(P-q)^2} + \frac{i\,n_d}{m_0^2q^2(q-k)^2} \right.\right. +\nl
&&\qquad\qquad\quad +\; \left.\left. \frac{i\,n_e}{m_0^2q^2(P-q-k)^2} \right\rbrace  + \frac{\bar n_c}{g^2} \left(\delta Z_g + \frac{3}{2}\delta Z_3\right) - \frac{\bar n_d + \bar n_e}{g^2}\delta Z_3   \right]  \nl
&+& \frac{P^2}{2\pi} \int\frac{d^Dq}{(2\pi)^D} (2\pi)\delta\left[(q-k)^2\right] (2\pi)\delta\left[(P-q)^2\right] \;\times\nl
&&\qquad\qquad\qquad \times\; \left\lbrace \frac{n_f}{q^4} + \frac{n_g}{(P-k)^4} + \frac{n_h}{ q^2 (P-k)^2} + \frac{n_i}{q^2 (P-q+k)^2} \right\rbrace \;.\nl
\ea{evol4}
Terms in $A(z,P^2)$ are calculated in Appendix~\ref{sec:calcSF}.

In order to keep only the LL terms, and to eliminate the collinear divergence from $A(z,P^2)$, we differentiate \REF{evol4} with respect to $t=\ln P^2$ to obtain 
\be
\partial_t \mathcal D\left(x,P^2\right) \;=\; \frac{g^4}{2}\int\frac{dz}{z}\, d_0\left(\frac{x}{z},m_0^2\right) \partial_t A(z,P^2)
\ee{evol5}
for the dimensionless function $\mathcal D(x,P^2) = P^2 D(x,P^2)$. (Derivations of the coupling $g$ results in terms of $\mathcal O(g^3)$, and thus are neglected.) As $\mathcal D = \dfrac{g^2}{2} d_0 + \mathcal O(g^4)$, we arrive at the DGLAP equation:
\be
\partial_t \mathcal D\left(x,P^2\right) \;=\; g^2\int\frac{dz}{z}\,\mathcal D\left(\frac{x}{z},P^2\right) \Pi(z,P^2)\;,
\ee{evol6}
with splitting function (SF)
\be
\Pi(z) \;=\; \frac{\partial}{\partial\ln P^2}A(z,P^2) \;=\; \frac{n_f}{(4\pi)^3}\frac{1-z}{z^2}  - \frac{n_c}{2 (4\pi)^3} \delta(1-z) \;.
\ee{evol7}
Note that the SF is proportional to the distribution of daughter partons in the LL approximation at $\mathcal{O}(g^4)$: $p_0{\dfrac{dN}{d^5p}}^{DP}\sim \Pi(z,P^2)$, thus, ${\dfrac{dN}{dz}}^{DP} \sim z^3\Pi(z) = a z(1-z)+b\,\delta(1-z)$, which is of the form of the SF used in \cite{bib:Graz}.

\subsection{Solving the DGLAP equation}
\label{sec:DGLAPsolv}
Introducing Mellin transforms $\tilde f(\omega) = \int\limits_0^1 dx x^{\omega-1} f(x)$, the DGLAP equation \REF{evol6} simplifies to
\be
\partial_t \tilde{\mathcal D}\left(\omega,P^2\right) \;=\; g^2\,\tilde{\mathcal D}\left(\omega,P^2\right) \tilde \Pi(\omega)\;,
\ee{evol8}
where the Mellin-transform of the SF is
\be
\tilde\Pi(\omega) \;=\; \frac{1}{(4\pi)^3}\left[ \frac{n_f}{(\omega-1)(\omega-2) }  - \frac{n_c}{2}\right] \;.
\ee{evol9}
The solution of the DGLAP equation is
\ba
\tilde{\mathcal D}\left(\omega,P^2\right) &\;=\;& e^{b(P^2)\tilde \Pi(\omega)}\tilde{\mathcal D}\left(\omega,P_0^2\right),\quad \textrm{with}\quad b(P^2) = \int\limits_{\ln P^2_0}^{\ln P^2} dt\, g^2 = \frac{2}{\beta_0}\ln\left[\frac{\ln(P^2/\Lambda^2)}{\ln(P_0^2/\Lambda^2)} \right] \nl
&& \quad \textrm{and coupling}\quad g^2 =\frac{2}{\beta_0\ln(P^2/\Lambda^2)} \;. 
\ea{evol10}

If we substitute our form of the FF in \REF{evol2} (and take into account that $\mathcal D(x,P^2) = P^2 D(x,P^2)$), we obtain that
\be
M^{12-D} \rho(M^2) \tilde{d}(\omega, M^2) = e^{b(P^2)\tilde \Pi(\omega)} M_0^{12-D} \rho(M_0^2) \tilde{d}(\omega, M_0^2) \;, 
\ee{evol11}
where $d(x,M_0^2)$ is the hadron distribution at a low initial scale in \REF{mic5}, for which, we have constructed the statistical model in Sec.~\ref{sec:D0}. To solve for both the jet mass distribution $\rho(M^2)$ and the hadron distribution $d(x,M^2)$, we exploit the normalisation condition 
\be
\bar n(M^2) = \int\frac{d^{D-1}p}{p} d\left(\frac{2p}{M}, M^2\right) =   \left(\frac{M}{2}\right)^{\omega^\ast}\kappa_{D-1} \tilde{d}(\omega^\ast, M^2) \;, 
\ee{evol12}
with $\omega^\ast=D-2$ and solid angle $\kappa_D = 2\pi^{D/2}/\Gamma(D/2)$. Consequently, when taking \REF{evol11} at $\omega=\omega^\ast$, $\tilde d$ drops out, and we get the solution for $\rho$:
\be
\rho(M^2) = \left(\frac{M_0}{M}\right)^{14-2D} \frac{\bar n_0}{\bar n} \rho(M_0^2) e^{b(P^2)\tilde \Pi(\omega^\ast)}   \;. 
\ee{evol13}
Writing this back into \REF{evol11}, we obtain the hadron distribution in the jet:
\be
\tilde{d}(\omega, M^2) = e^{b(M^2)[\tilde \Pi(\omega) - \tilde \Pi(\omega^\ast)]} \left(\frac{M_0}{M}\right)^{D-2} \frac{\bar n}{\bar n_0}\, \tilde{d}(\omega, M_0^2)   \;. 
\ee{evol14}
Now, we can readily obtain the FF by substituting \REF{evol13} and \REF{evol14} into \REF{evol2}: 
\be
\tilde{D}(\omega, M^2) = M_0^{10-D}\left(\frac{M_0}{M}\right)^{2} \rho(M_0^2)\, \tilde{d}(\omega, M_0^2)\, e^{b(M^2)\tilde \Pi(\omega)}   \;. 
\ee{evol15}
As we have a model for $d(x,M_0^2)$, we could use the inverse Mellin-transform to calculate the FF, however, we will use a less complicated approximation in the next section. 

Finally, the $p-$integrated cut blobs being the factors, which jets contribute to cross sections, if we do not measure the hadrons in them, take the form 
\ba
\bar{D}(M^2) &=& \int \frac{d^{D-1}p}{p} D\left(\frac{2p}{M}, M^2\right) = \kappa_{D-1} \left(\frac{M}{2} \right)^{D-2} D(\omega^\ast,M^2)\nl
&=& M^{D-4}M_0^{14-2D} \rho(M_0^2)\,  e^{b(M^2)\tilde \Pi(\omega^\ast)} \sim M^{D-4} \ln^{2\tilde\Pi(\omega^\ast)/\beta_0}(M^2/\Lambda^2)   \;. 
\ea{evol17}

\subsection{Scale evolution of the $q$ and $T$ parameters of the TS distribution}
\label{sec:TSevol}
Although the TS function \REF{mic6} is not an exact solution of the DGLAP equation, along with the hadron multiplicity distribution \REF{mic4}, it describes measured data of hadrons stemming from jets in $e^+e^-$ \cite{bib:UKee} and $pp$ \cite{bib:UKpp,bib:UKpp3D} collisions at various energy scales. Based on this, we conjecture that \REF{mic5} is a reasonably good approximation of $d(x,M^2)$, the hadron distribution in a jet at any scale $M$. To determine the scale dependence of the parameters $\bar n$ and $\sigma$ of the model, let us insert $d(\omega,M) = \sum\limits_n \mathcal{P}_n(M)\,n\,\tilde d_n(\omega,M)$ 
with 
\be
\tilde d_n(\omega,M) \;=\;   \frac{\left(\frac{2}{M} \right)^{\omega^\ast} \Gamma(\omega) \Gamma\left(\frac{\omega^\ast n}{2}\right)}{\kappa_{D-1} \Gamma(\omega^\ast) \Gamma\left(\omega + \frac{\omega^\ast (n-2)}{2}\right)} 
\ee{TS1}
into \REF{evol14}, to obtain 
\be
\sum_n \mathcal{P}_n(M)\,n\,\tilde d_n(\omega,M) = e^{b(M^2)[\tilde \Pi(\omega) - \tilde \Pi(\omega^\ast)]} \left(\frac{M_0}{M}\right)^{\omega^\ast} \frac{\bar n}{\bar n_0}\, \sum_n \mathcal{P}_n(M_0)\,n\,\tilde d_n(\omega,M_0) \;.
\ee{TS2}
As the TS distribution is only an approximation of the exact solution of the DGLAP equation, we cannot eliminate $\omega$ from this equation. Nevertheless, we may require \REF{TS2} to hold for a suitable set of fixed values of $\omega$ to force the scale evolution of certain moments of $x$ to be exact. In this paper, we prescribe that \REF{TS2} hold for $\omega = \lbrace \omega^\ast,\omega^\ast\pm1\rbrace$ (which refers to momenta $\left\langle\frac{1}{x}\right\rangle,\langle1\rangle$, and $\langle x\rangle$) and obtain
\ba
\bar n(M) &\;=\;& \bar n_0\, e^{-b\, a_+} \;,\nl
\sigma^2(M)   &\;=\;& \bar n \left[\frac{2}{\omega^\ast} + e^{b\, a_-}\left( \frac{\sigma_0^2}{\bar n_0} + \bar n_0 - \frac{2}{\omega^\ast}\right) \right] - \bar n^2 \;,
\ea{TS3}
with $b=b(M^2)$, $a_{\pm} =  \tilde\Pi(\omega^\ast\pm1) - \tilde\Pi(\omega^\ast)$ and initial values $\sigma(M_0) = \sigma_0, \bar n(M_0) = \bar n_0$. Furthermore, we choose the initial scale $M_0$ to be the scale, at which, the multiplicity distribution becomes Poissonian ($\sigma^2_0 = \bar n_0$), to get even simpler results:
\ba
\bar n(M) &\;=\;& \bar n_0\, e^{-b\, a_+} \;,\nl
\sigma^2(M)   &\;=\;& \bar n \left[\frac{2}{\omega^\ast} + e^{b\, a_-}\left( 1 + \bar n_0 - \frac{2}{\omega^\ast}\right) \right] - \bar n^2 \;.
\ea{TS4}
The obtained logarithmic growth of the mean hadron multiplicity with the energy scale $\bar n\propto \ln^a(M)$ is in accordance with observations. Consequently, the scale dependence of the standard parameters of the hadron multiplicity distribution is
\ba
p &\;=\;& 1 - \frac{\bar n}{\sigma^2} = 1 - \frac{1}{\frac{2}{\omega^\ast} + e^{b\, a_-}\left( 1 + \bar n_0 - \frac{2}{\omega^\ast}\right) - \bar n_0\, e^{-b\, a_+}}\;,\nl
r &\;=\;& \frac{\bar n^2}{\sigma^2 - \bar n} = \frac{\bar n_0\, e^{-b\, a_+}}{\frac{2}{\omega^\ast} + e^{b\, a_-}\left( 1 + \bar n_0 - \frac{2}{\omega^\ast}\right) - \bar n_0\, e^{-b\, a_+} - 1 } \;.
\ea{TS5}
In $D=4$ dimensions, the FF takes the form of the TS distribution \REF{mic6}, and the scale evolution of its parameters (using \REF{mic7}) is
\ba
q - 1&\;=\;& \frac{\sigma^2 - \bar n}{\bar n^2 + 3(\sigma^2 - \bar n)} = \frac{1 - e^{-b\,(a_+ + a_-)}}{3 - 2 e^{-b\, (a_+ + a_-)}}\;,\nl
\tau &\;=\;& \frac{\bar n}{\bar n^2 + 3(\sigma^2 - \bar n)} =  \frac{\tau_0 }{3\,e^{b\, a_-} - 2\, e^{-b\, a_+}}\;,
\ea{TS6}
having introduced $\tau_0 = 1/n_0$ in accordance with the equipartition principle. This type of logarithmically rising tendency of $q$, and falling tendency of $\tau$ has been observed in various high-energy collisions listed in Sec.~\ref{sec:intro}. Looking at the asymptotic behaviour, it is interesting to point out that $q$ is bounded from above: $q \; \rightarrow \;  \frac{4}{3}$.

Since now we have the approximate form of $\tilde d(\omega,M^2)$, we may use it in the expression of the FF $\tilde D(\omega,M^2)$ via only substituting \REF{evol13} into \REF{evol2}: 
\be
\tilde{D}(\omega, M^2) = M^{10-D} \left(\frac{M_0}{M}\right)^{14-2D} \frac{\bar n_0}{\bar n}\rho(M_0^2)\, \tilde{d}(\omega, M^2)\, e^{b(M^2)\tilde \Pi(\omega^\ast)} \sim \tilde{d}(\omega, M^2)  \;. 
\ee{TS7}
From the point of view of the $\omega$ dependence, this formula differs from \REF{evol15} in the change of $\tilde{d}(\omega, M_0^2)\, e^{b(M^2)\tilde \Pi(\omega)} \rightarrow \tilde{d}(\omega, M^2)$.

\label{sec:ee2hX}
\begin{figure}[!h]
\begin{center}
\includegraphics[width=0.8\textheight]{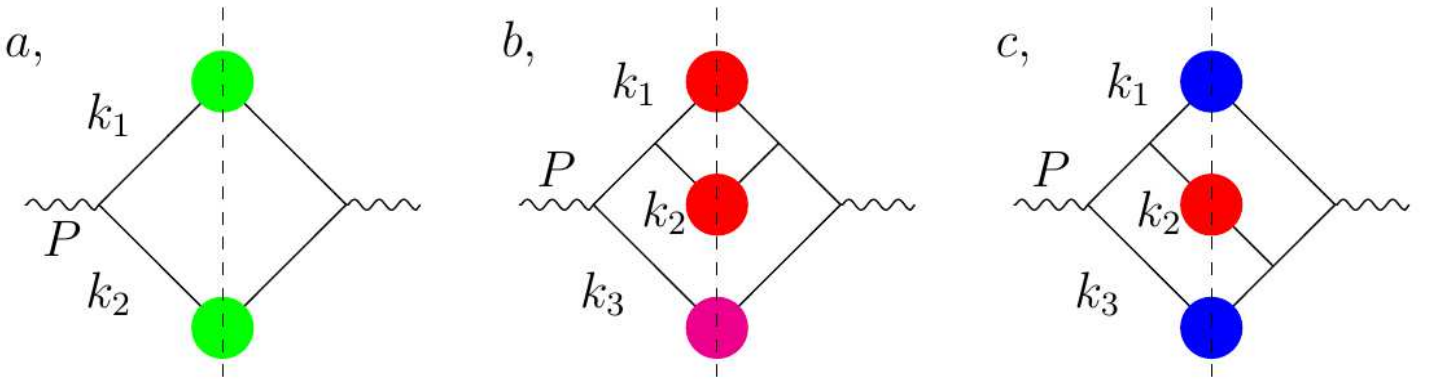} %PLB
\end{center}
\caption{The partonic part of 2--, and 3--jet final states in the $e^+e^-\rightarrow hX$ process. We will refer to the process shown in \textbf{panel b} as the \textbf{\textit{``split''}}, while the one in \textbf{panel c} the \textbf{\textit{``crossed''}} 3-jet event.
\label{fig:3j}}
\end{figure}
\begin{figure}[!h]
\begin{center}
\includegraphics[width=0.6\textheight]{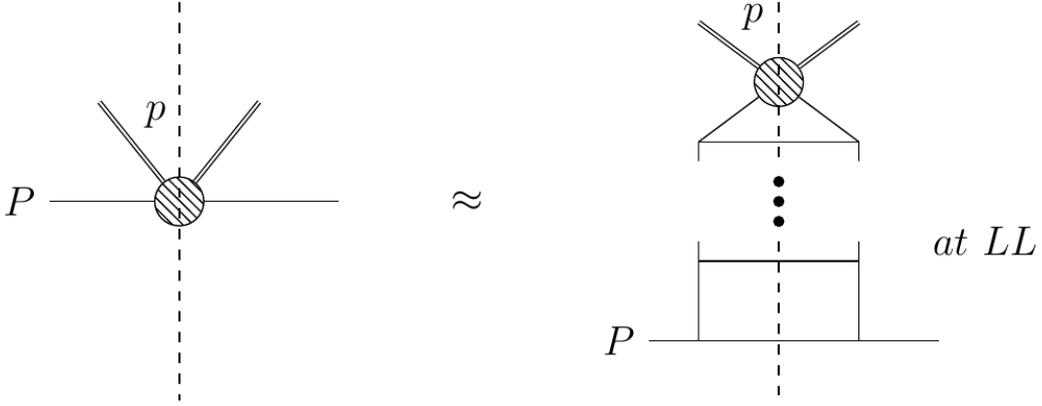} %PLB
\end{center}
\caption{The cut blob in Fig.~\ref{fig:3j} denoting the fragmentation pocess of a virtual leading parton of momentum $P$ fragmenting into hadrons, among which, one has momentum $p$ in the LL approximation.
\label{fig:1j}}
\end{figure}

\section{Hadron and jet distributions in $e^+e^-\rightarrow2-3$ jet events}
\label{sec:res}
Similarly to \cite{bib:Graz,bib:CollinsMC}, we mimic the $e^+e^-\rightarrow\gamma^\ast\rightarrow q\bar q$ process via introducing the term $eA\phi^2$ into the Lagrangian, thus, coupling a 'scalar photon' to the `scalar parton' field of the $\phi^3$ model. When calculating the distributions of jets stemming from $e^+e^-$ annihilations with 2-, and 3-jet final states depicted in Fig.~\ref{fig:3j}, we use the un-cut propagators of the leading partons attached to the $\bar D(k_i)$ cut blobs given in \REF{evol17}. These cut blobs are the resummed LL terms of the process (shown in Fig.~\ref{fig:1j}) of a virtual leading parton emitting on-shell daughter partons before fragmenting into hadrons when its virtuality reaches some low scale $M_0$. As the cut blobs along with the parton legs of momenta $k_i$ on both their sides contribute a factor of $\bar D (k_i^2)/k_i^4$, we introduce the function 
\be
F_n(k_1,\dots,k_n) \;=\; \prod_{i=1}^n \frac{\bar D(k_i^2)}{k_i^4} \,\delta^D\left(\sum_{j=1}^n k_j - P\right) \propto \prod_{i} \frac{\ln ^\alpha\left(\dfrac{k_i^2}{\Lambda^2}\right) }{k_i^{8-D}}  \,\delta^D\left(\sum_j k_j - P\right) \;,
\ee{jet1}
with $\alpha = 2\tilde\Pi(\omega^\ast) / \beta_0$, and $P=(\sqrt s,\mathbf{0})$ being the total momentum of the incoming $e^+e^-$ pair. This way, the momentum distribution of jets in a 2--jet event reads
\be
\frac{1}{\sigma_{2jet}}\frac{d\sigma_{2jet}}{d^Dk_1 d^Dk_2} \;=\; F_2(k_1,k_2)\;,
\ee{jet2}
and we can make out the 3--jet cases in panels \textit{\textbf{b}} and \textit{\textbf{c}} of Fig.~\ref{fig:3j} as  
\be
\frac{1}{\sigma^b_{3jet}}\frac{d\sigma_{3jet}^b}{d^Dk_1 \dots d^Dk_3} = g^2 \frac{F_3(k_1,k_2,k_3)}{(k_1+k_2)^4} \;,\qquad
\frac{1}{\sigma^c_{3jet}}\frac{d\sigma_{3jet}^c}{d^Dk_1 \dots d^Dk_3} = g^2 \frac{F_3(k_1,k_2,k_3)}{(k_1+k_2)^2(k_2+k_3)^2} \;.
\ee{jet3}
As we do not focus on the renormalisation of the electric charge $e$, we have omitted the term with the radiative correction to the photon-parton vertex in the 2--jet case in Fig.~\ref{fig:3j}.a. 

In order to regularize the divergences originating from the poles of the propagators of leading partons, we prescribe an extra condition $k_i^2\geq m_0^2$ so that jet masses be larger than some low energy scale $m_0>\Lambda$, being of the order of the proton mass. Such a requirement is natural, as the jet mass cannot be smaller than the sum of the masses of the hadrons it containes. This condition also eliminates the collinear divergence from the 3-jet events, which would arise if leading partons were taken to be on-shell. In models relying on factorisation, such divergences may be removed, for example, by the subtraction of the configurations, in which, some jet momenta are parallel \cite{bib:CollinsMC}.

When calculating distributions of quantities involving jet momenta numerically, we first generated jet momenta randomly according to the uniform distribution in the phasespace, and accepted them, if they satisfied the condition of $k_i^2\geq m_0^2$. Then, we filled the calculated quantities of interest into histograms weighted by the corresponding jet cross-sections in \REF{jet2} and \REF{jet3}. To generate random jet momenta, we exploited that a thermal ensemble readily provides particles with uniform momentum distribution in the phasespace. In fact, given $n$ particles, having an arbitrary set of initial momenta $k_i$ ($\sum k_i = P$), after at least $10\,n$ pairwise \textit{``collisions''}, they reach a thermal state. In these imaginary collisions, the incoming particles exchange a random momentum, thus, the new momentum of one of the pair has uniform distribution in the center-of-mass (CM) frame:
\be
d\mathcal{P}(k) \;=\; dk^0\,d |\mathbf k|d\Omega\, |\mathbf{k}|^{D-2}\, \Theta(k_0 - |\mathbf{k}|)\, \Theta\left(E^{CM}-|\mathbf{k}| - k_0\right)\, \Theta\left(\frac{E^{CM}}{2} - |\mathbf k| \right)   \;.
\ee{jet4}
This method works also for a mixture of particles of various masses, in case of which, analytic formulas (like \REF{mic2}, we have obtained for massless, on-shell particles) for the direct generation of particle momenta one-by-one are not available. 

\begin{figure}%[!h]
\begin{center}
\includegraphics[width=0.6\textheight]{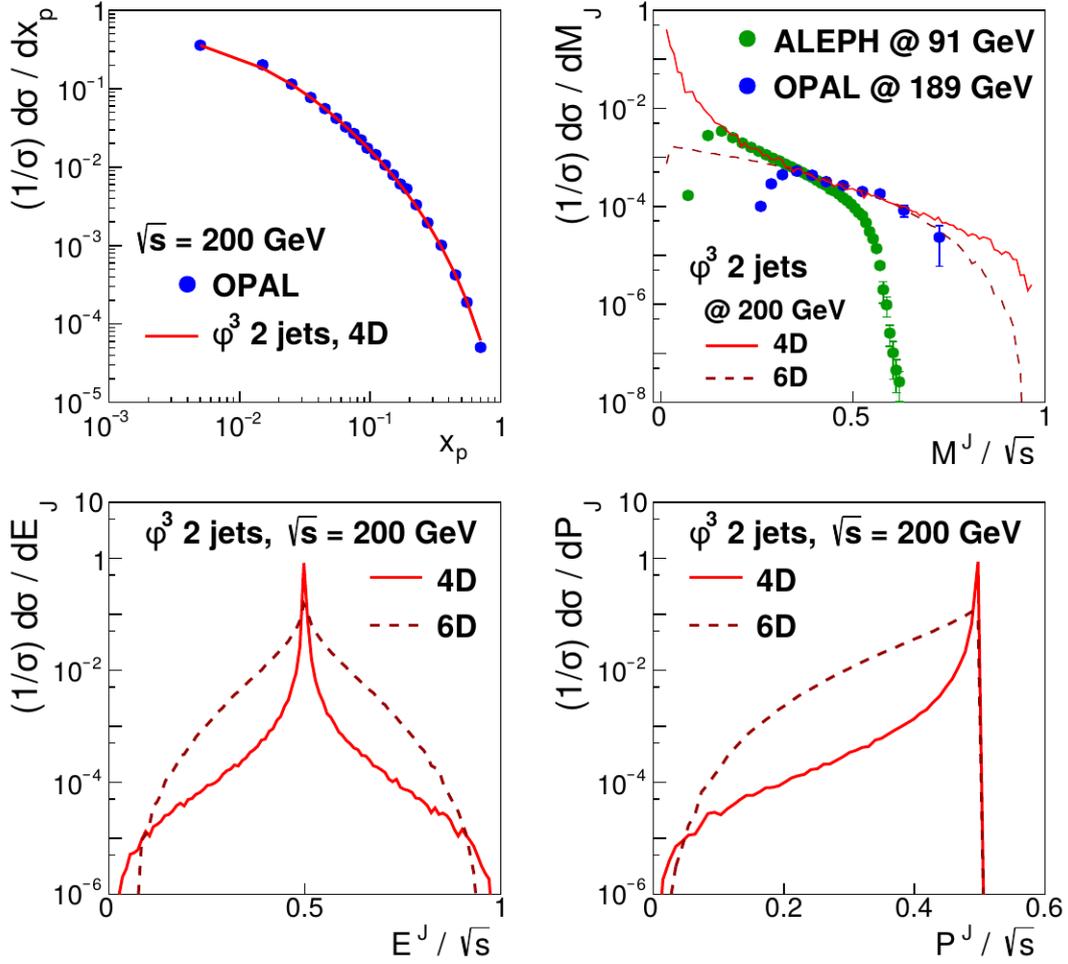} %PLB
\end{center}
\caption{Measured distributions of hadron energy (\textbf{top-left}) and heavy jet mass (\textbf{top-right}) in $e^+e^-$ annihilations compared with our calculations for 2-jet events based on the $\phi^3$ theory with virtual leading partons. \textbf{Bottom,} our results for jet energy and momentum distributions in 4 and 6 dimensions.
\label{fig:4Dcalib}}
\end{figure}

Having obtained the jet momenta $k_i$, the probability of a hadron stemming from the $i^{th}$ jet to have momentum $p$ is given by $d\left(\dfrac{2pk_i}{k_i^2},k_i^2\right)$ according to \REF{TS7}. The functional form of $d$ is given in \REF{mic5} (or \REF{mic6} in $D=4$ dimensions), and the values of its parameters depend on the jet mass according to \REF{TS4}--(\ref{TS6}). This way, the spectrum of hadrons stemming from the $i^{th}$ jet in an $n-$jet event is calculated as
\be
\frac{1}{\sigma^{h}_{n-jet}}p_0\frac{d\sigma^{h}_{n-jet}}{d^{D-1}\mathbf{p}} \;=\; \int\prod\limits_{j=1}^n d^Dk_j\, \frac{1}{\sigma_{n-jet}}\frac{d\sigma_{n-jet}}{d^Dk_1 \dots d^Dk_n} \, d\left(\dfrac{2pk_i}{k_i^2},k_i^2\right)\;.
\ee{jet5}
Setting the $M_0$ starting scale of the DGLAP evolution of the FF to be equal to the lower cut-off for the allowed jet masses, $M_0=m_0$, our model has 4 parameters: the mean hadron multiplicity $\bar n_0$ at starting scale $M_0$, along with the $\beta_0$ and $\Lambda$ parameters of the coupling. We calibrated these parameters via fitting the calculated hadron distribution in 2-jet events in $D=4$ dimensions to the spectrum of charged hadrons measured at $\sqrt s=$ 200 GeV by the OPAL Collaboration \cite{bib:MjetEE1}, and obtained the following values: $\Lambda$ = 98.3 MeV, $\beta_0$ = 0.08107, $\bar n_0$ = 6.814, $M_0$ = 3.038 GeV/$c^2$. Fit results are shown in the \textbf{top-left panel of Fig. \ref{fig:4Dcalib}}. As the $\phi^3$ model is renormalizable in $D=6$ dimensions, we used the formulas for the coupling and the SF $\tilde\Pi(\omega^\ast)$ obtained in 6 dimensions, even when calculating the 4-dimensional results.  This inaccuracy is, however, of not much relevance as the aim of this paper is to derive the scale dependence of the parameters of the TS distribution in a high-energy process, and to show the effect of using virtual leading partons even in the ``hard'' part of cross sections. 

In the \textbf{top-right panel of Fig. \ref{fig:4Dcalib}}, we compare our results on the distribution of the mass of jets in 2-jet events in 4 and 6 dimensions with the available experimental result, which is the distribution of the heavy jet mass (the larger among the two total momentum squared of particles in the semispheres of the phasespace). Although, OPAL data were taken at a slightly lower collision energy ($\sqrt s=189$ GeV), our results are in accordance with it in the high-jet mass range. For lower values, the heavy jet mass distribution decreases rapidly, and we do not expect our result to describe it correctly. Out of curiosity, we have also plotted a dataset measured at a lower energy $\sqrt s=$ 91 GeV.

In the \textbf{bottom panels of Fig.~\ref{fig:4Dcalib}}, we compare the calculated distributions of the energy and the vector part of jet momenta in 2-jet events in 4 and 6 dimensions. As expected, these distributions have a sharp peak at $\sqrt s/2$, where the distributions of on-shell particles would have a $\delta$-peak. Besides, the 6-dimensional distributions are broader, due to the lower power of jet virtualities in the denominator of \REF{jet1}.

\begin{figure}[!h]
\begin{center}
\includegraphics[width=0.6\textheight]{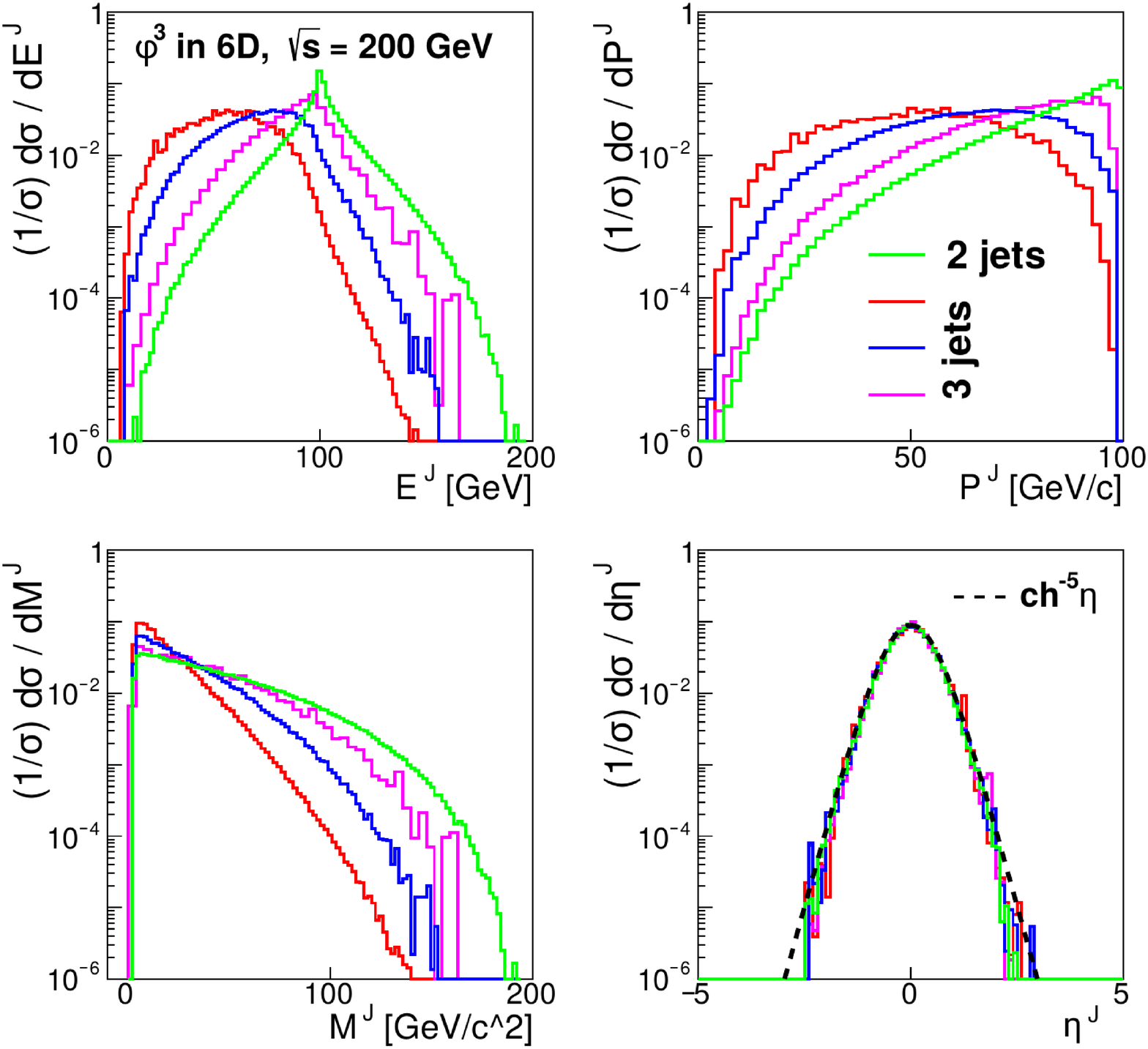} %PLB
\end{center}
\caption{Energy, momentum, mass and rapidity distributions of jets in 2--, and 3--jet final states stemming from $e^+e^-$ annihilations calculated at $\sqrt s=$ 200 GeV in $D=6$ dimensions. Colors show which graph refers to which jet in Fig.~\ref{fig:3j}.
\label{fig:jetDists}}
\end{figure}
\textbf{Fig.~\ref{fig:jetDists}} shows distributions of jet energies, momenta, masses and rapidities ($\eta^J = \ln\left(\frac{E^J+P^J_z}{E^J-P^J_z}\right)$, where $P^J_z$ is the component of $P^J$, which is parallel to the beam axis) calculated in 6 dimensions. Color encoding shows which histogram refers to which jet in Fig.~\ref{fig:3j}, where the Feynmann graphs of the corresponding 2-, and 3-jet processes are depicted. It is trivial that distributions of the first two jets of momenta $k_1$ and $k_2$, colored in red in Fig.~\ref{fig:3j}.b (which we will refer to as the \textbf{``split''} process) coincide. So do the distributions of the first and third jets of momenta $k_1$ and $k_3$, colored in blue in Fig.~\ref{fig:3j}.c (which we will call the \textbf{``crossed''} process). However, it is interesting that the second jet in the \textit{crossed} process has the same distributions as do the first two jets in the \textit{split} process. Besides, a clear hierarchy among the mean jet energies, momenta, and masses are visible. This seems to be in connection with which \textit{``generation''} the leading parton of a jet is produced. In Fig.~\ref{fig:3j}, we see that green jets belong to the $1^{st}$ generation, as their leading partons were created in the first splitting. Consequently, their average energy $\bar E^J$ is the largest. The magenta jet also belongs to the $1^{st}$ generation, but it stems from a 3-jet event, where the total energy is distributed among more  jets then in the 2-jet case, thus, the $\bar E^J$ of the magenta jet is somewhat smaller. The $\bar E^J$ of red jets is the smallest, because they belong to the $2^{nd}$ generation, whereas the $\bar E^J$ of the blue jets is in between those of the red and magenta jets, as the blue jets come from the interference of a $1^{st}-$ and a $2^{nd}-$generation jet.

Rapidity, as well as, anglular distributions are the same for all the jets due to rotational symmetry.

\begin{figure}[!h]
\begin{center}
\includegraphics[width=0.33\textwidth]{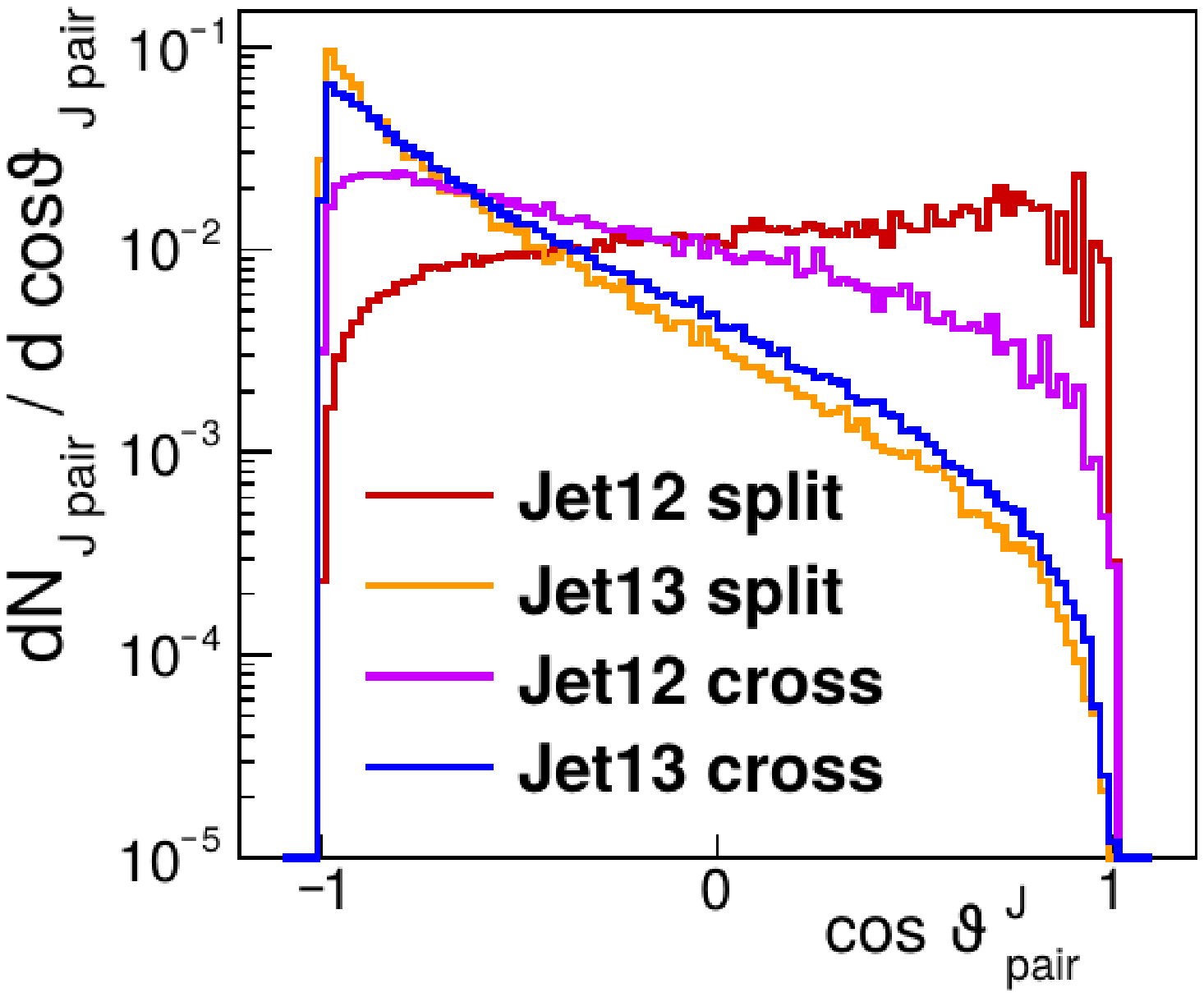}
\includegraphics[width=0.63\textwidth]{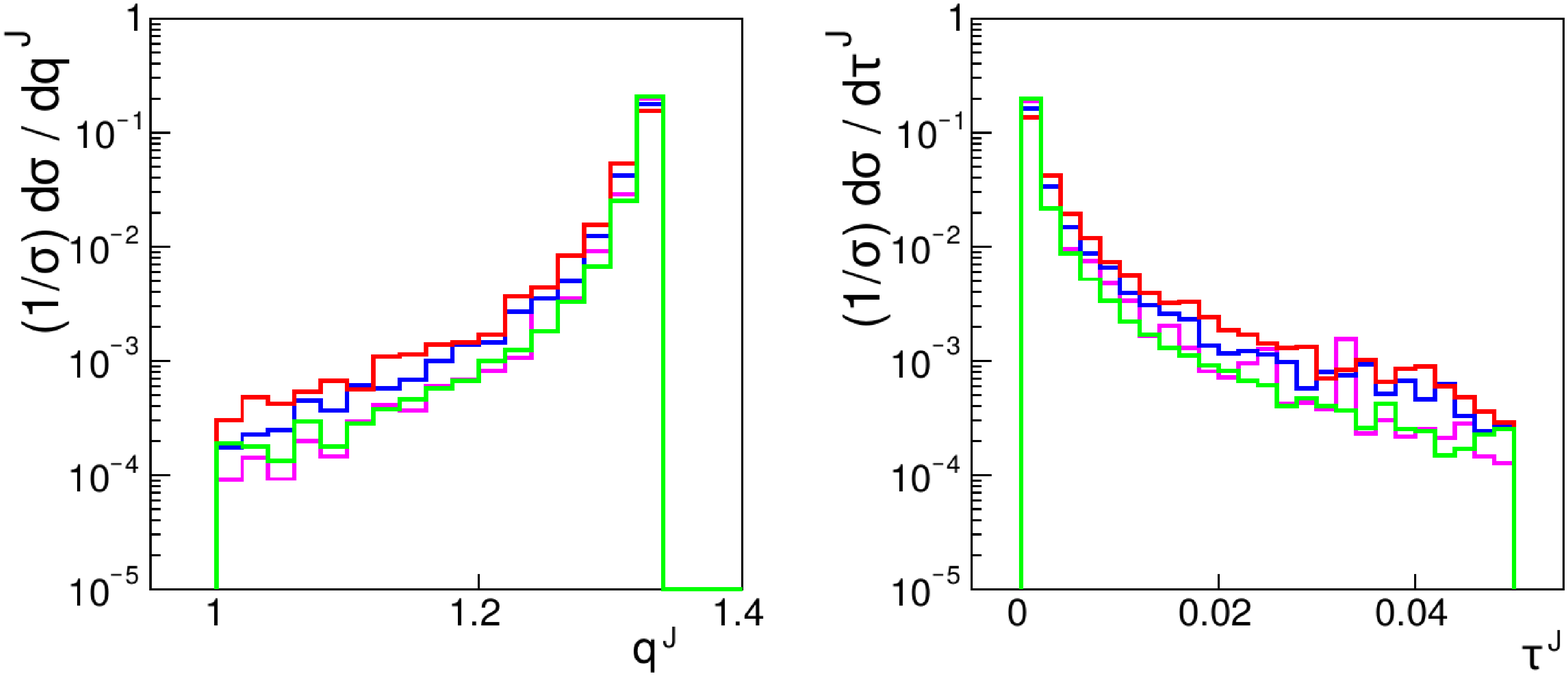} %PLB
\end{center}
\caption{Distributions of the angles between jet momenta (\textbf{left}). Distribution of the $q$ (\textbf{middle}) and $\tau$ (\textbf{right}) parameters of the spectrum of hadrons in jets. Colors show which graph refers to which jet in Fig.~\ref{fig:3j}.
\label{fig:qT}}
\end{figure}
From the \textbf{left panel of Fig.~\ref{fig:qT}}, we can see that in a \textit{split} process, the angle between the red jets $\theta_{12}$ and the one between the red and the magenta jet $\theta_{13}$ are most likly to be ordered as $\theta_{12}<\dfrac{\pi}{2}<\theta_{13}$, whereas  jets in a \textit{crossed} process favour the three-pronged star configuration.

As jet masses fluctuate event-by-event in $e^+e^-$ annihilations, so do the $q$ and $\tau$ parameters of hadron spectra in jets due to \REF{TS6}. According to the \textbf{middle and left panels of Fig.~\ref{fig:qT}}, $q$ and $\tau$ has sharp distributions. Consequently, the TS distribution provides a good description of the hadron spectrum, which is the sum of spectra in single jets averaged over jet-momentum fluctuations. It is important to point out, that the scale evolution of FFs can be examined even if the measured spectrum is available at only a single value of $\sqrt s$, if we set the jet mass as fragmentation scale.

\begin{figure}[!h]
\begin{center}
\includegraphics[width=0.8\textheight]{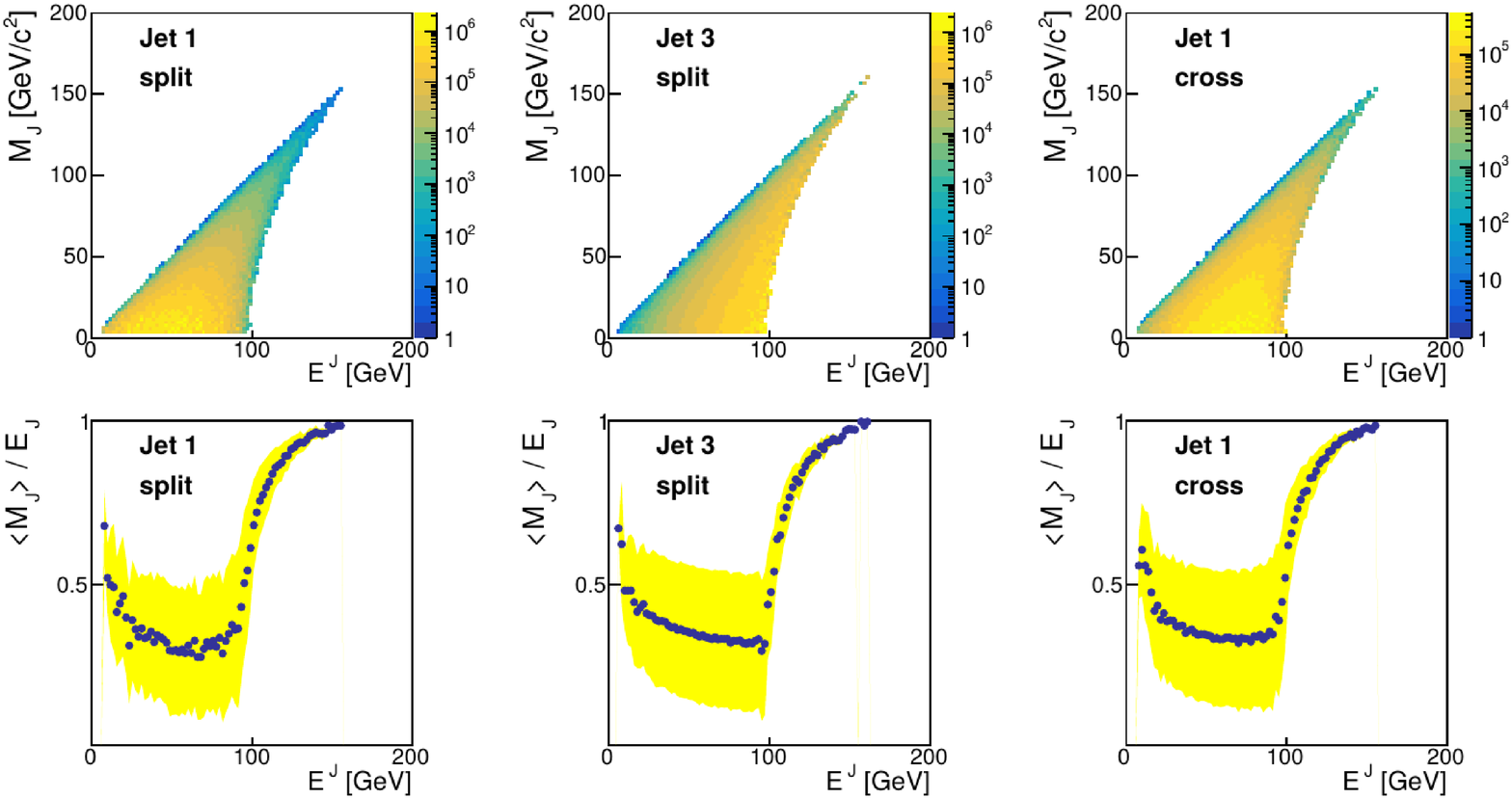} %PLB
\end{center}
\caption{\textbf{Top,} double differential distributions of the jet mass and energy. \textbf{Bottom,} average jet mass vs. jet energy. Legend shows which graph refers to which jet in the \textit{split} and \textit{crossed} events in Fig.~\ref{fig:3j}.
\label{fig:jetEM}}
\end{figure}
As factorisation theorems are expected to work best when jet masses are small compared to jet energies, we have plotted the double-differential distribution $d\sigma^J/dE^JdM^J$ along with the dependence of the $\langle M^J\rangle/E^J$ ratio as a function $E^J$ in \textbf{Fig.~\ref{fig:jetEM}} for the three types of jets produced in our model in 3-jet events. As it can be seen, although, the most probable events are those with small jet masses, the average jet mass is never negligable compared to the jet energy. When one of the jets acquires energy $E^J\geq\sqrt s/2$, the situation gets even worse, since in that case, the energies as well as the momenta of the other jets are small. Consequently, all jet momenta has to be small, thus, the mass of the jet with large $E^J$ has to be large as well.   

\begin{figure}[!h]
\begin{center}
\includegraphics[width=0.8\textheight]{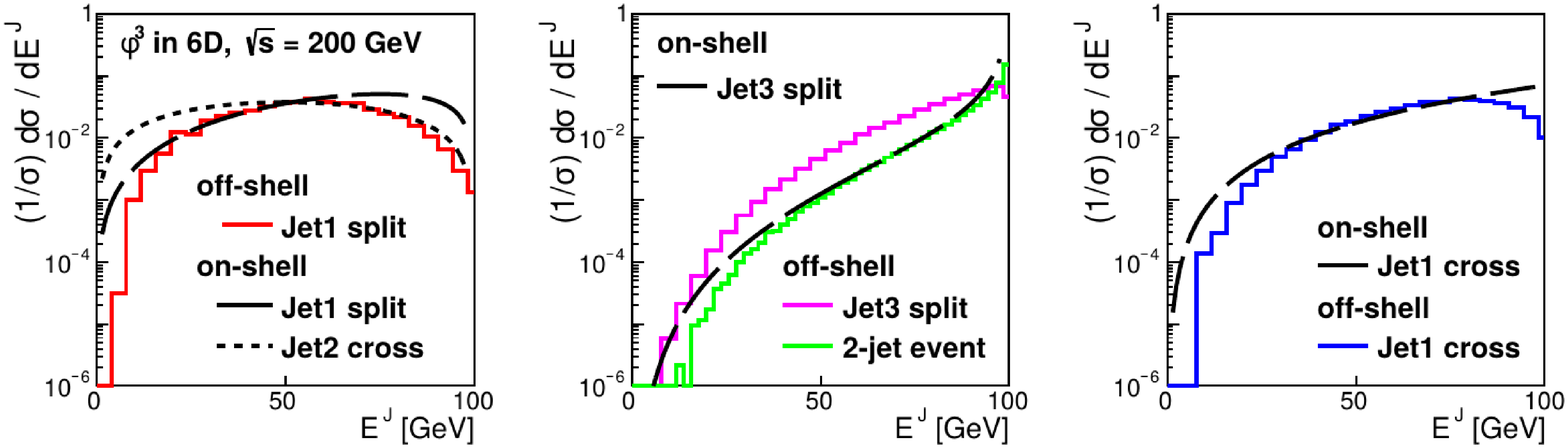} %PLB
\end{center}
\caption{On- vs. off-shell jet energy distributions. Colors show which graph refers to which jet in Fig.~\ref{fig:3j}.
\label{fig:onOff}}
\end{figure}
At this point, it is not surprising that jet energy distributions of our model (using off-shell leading partons and LL resummation for the fragmentation process) differ significantly from those, obtained using fixed-order cross sections with on-shell leading partons, as  can be seen in \textbf{Fig.~\ref{fig:onOff}}. The distribution of the on-shell leading partons in 3-jet events are 
\ba
{\frac{d\sigma}{dx}}^{on-shell} &\;\sim\;& x(1-x) [1 - A\ln(1-x)] \quad \textrm{for jet 1-2 and}\nl
&\;\sim\;& \frac{x^3}{1-x} \quad \textrm{for jet 3 in a \textit{split} event;}\nl
&\;\sim\;& x^2 \quad \textrm{for jet 1-3 and}\nl
&\sim& x(1-x)  \quad \textrm{for jet 2 in a \textit{crossed} event}.
\ea{jet6}
The Feynmann-graphs of the processes of the creation of these partons are depicted in Fig.~\ref{fig:DSE}.f--i (if we replace the jet blobs by cut free propagators). Consequently, the corresponding distributions are the terms, listed in \REF{1p31} multiplied by the phasespace factor $x^3$ (which we have already calculated when determining the SF). In the off-shell case, the distribution of jet 1-2 in a \textit{split} event coincides with that of jet 2 in a \textit{crossed} event. In the on-shell case, these distributions in \REF{jet6} only differ in the $\ln(1-x)$ term coming from the dimensional regularisation of the collinear divergence resulting from the configuration when jet 1 and 2 are parallel. A different regulatisation method might remove this term. According to Figs.~\ref{fig:jetEM}--\ref{fig:onOff}, the shape of the jet energy distributions obatained using on-, and off-shell partons are close to each other only within limited intervals around $E^J\approx\sqrt s/4\pm\sqrt s/8$, where $\langle M^J\rangle/E^J\approx0.3$. Outside of this region, the $\langle M^J\rangle/E^J$ ratio is much higher, and the on-, and off-shell distributions differ significantly. 

\begin{figure}[!h]
\begin{center}
\includegraphics[width=0.45\textheight]{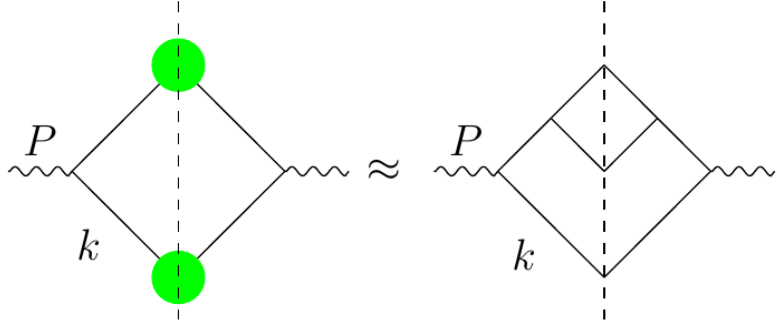} %PLB
\end{center}
\caption{\textbf{Left,} 2-jet event with off-shell leading partons, where the blobs denote the fragmentation process resummed in the LL approximation. \textbf{Right,} 3-jet event with on-shell leading partons.
\label{fig:onOff2}}
\end{figure}
In the \textbf{middle panel of Fig.~\ref{fig:onOff}}, we can see that, although, the shape of the on-shell distribution of jet 3 is not close to that of the off-shell one, it approximates the off-shell distribution in 2-jet events. This observation suggests that, having a pair of jets created in a splitting (as depicted in \textbf{Fig.~\ref{fig:onOff2}}), in order to get a good approximation for the energy distribution of the jet of momentum $k$, it is enough to keep only the virtuality of the leading parton of the other jet. Furthermore, it is also enough to keep only the first splitting in the other jet, instead of resumming the whole 'parton ladder' in the LL approximation. This result supports the argument in \cite{bib:CollinsMC} that, when we create a parton shower, if we neglect the virtualities of partons of a given generation due to using on-shell cross sections, our inaccuracy is compensated for when we create the next generation. 

\begin{figure}[!h]
\begin{center}
\includegraphics[width=0.6\textheight]{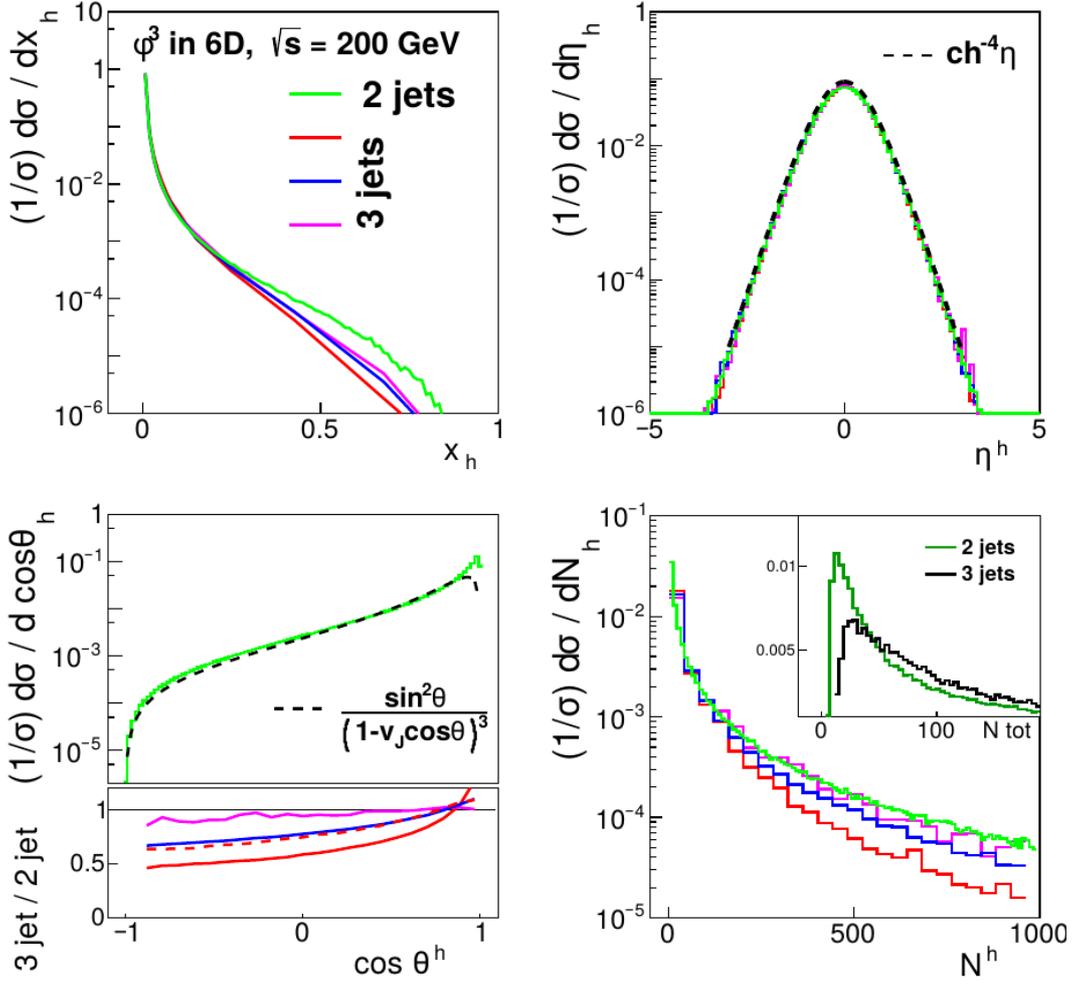} %PLB
\end{center}
\caption{Hadron energy (\textbf{top-left}), rapidity (\textbf{top-right}), angle (\textbf{bottom-left}) and multiplicity (\textbf{bottom-right}) distributions. Colors show which graph refers to which jet in Fig.~\ref{fig:3j}.
\label{fig:hadDists}}
\end{figure}
The spectrum, rapidity and angular distributions of hadrons stemming from each jet in 2-, and 3-jet events, shown in \textbf{Fig.~\ref{fig:hadDists}}, can be obtained using \REF{jet5}. Besides, the distribution of the $N_h$ number of hadrons in a given jet is  
\be
\frac{1}{\sigma^{h}_{n-jet}}\frac{d\sigma^{h}_{n-jet}}{dN_h} \;=\; \int\prod\limits_{j=1}^n d^Dk_j\, \frac{1}{\sigma_{n-jet}}\frac{d\sigma_{n-jet}}{d^Dk_1 \dots d^Dk_n} \, \mathcal{P}_{N_h}(k^2)\;.
\ee{jet7}
The difference between these distributions are due to the hierarchy of the average mass, energy and momentum of the different jets (see Fig.~\ref{fig:jetDists}) containing the hadrons. The more energetic the jet, the more high-energy hadrons stem from it, as can be seen in the \textbf{top-left panel}. The larger the jet mass, the wider the phasespace ellipsoid of the hadrons in the jet, thus, the average angle between hadron momenta and the momentum of the mother jet is larger as well, according to the \textbf{bottom-left panel}. The mean number of hadrons in a jet grows monotonically with the jet mass in accordance with the \textbf{bottom-right panel} and \REF{TS4}. Finally, the shape of hadronic pseudo-rapidity ($\eta = \ln\frac{p_0+p_z}{p_0-p_z}$) distributions shown in the \textbf{top-right panel} can be derived from rotational invariance and dimensional arguments.

\subsection{A micro-canonical event generator}
\label{sec:EvGen}
We obtained the hadron distributions in the previous section via generating all the hadrons in the jets using the micro-canonical model presented in Sec.~\ref{sec:D0}. That is, after having generated the jet momenta, we used the NBD multiplicity distribution in \REF{mic4} (the parameters of which depend on the jet mass according to \REF{TS4}) to obtain the numbers of hadrons in the jets. After that, we generated the momenta of hadrons in each jet according to the micro-canonical ensemble using the random collisional method, described in the paragraph before \REF{jet4}. As in our approximation, hadrons are on-shell and massless, we replaced \REF{jet4} with  
\be
d\mathcal{P}(p) \;=\; d |\mathbf p|d\Omega\, |\mathbf{p}|^{D-2}\, \Theta\left(\frac{E^{CM}}{2} - |\mathbf p| \right)  
\ee{EG1}
for the generation of the momentum of one of the outgoing particles in the CM frame in each imaginary collision. This way, the one-particle distribution in each jet of $n$ hadrons becames of the form of \REF{mic3}, and the $n$-averaged distribution becames \REF{mic5}.

Since the results presented in the previous section only involve $d(x,k^2)$, the single-hadron distribution in a jet averaged over multiplicity fluctuations (the scale evolution of which, we have derived from first principles), actually, we used the statistical model only to obtain the form of the FF at an initial scale $M_0\approx$ 3 GeV (which value we obtained from fitting the model to measured data). However, the micro-canonical fragmentation model could, in principle, be used to obtain multi-particle observables as well, since it is very simple to generate momenta of all the particles in all jets event-by-event within its scope. Although, this way, we would discard the microscopic dynamics of the branching process of parton production within jets, as all measurables are derived solely from the phasespace of particles in a statistical ensemble.

\section{Summary}
\label{sec:fact}

\begin{itemize}
 \item We have summarized the status of the application of the Tsallis (TS) distribution in theoretical and experimental high-energy physics.

 \item We have resummed the fragmentation processes of a virtual leading parton (initiating the jet) emitting on-shell daughter partons in the leading-log approximation (LLA) in the $\phi^3$ theory. We have found that the fragmentation scale is the virtuality of the leading parton, which is equal to the jet mass $M_J$, and calculated the $M_J$-dependence of the $q$ and $T$ parameters (\REF{TS4}) of a TS-shaped fragmentation function (FF).
 
 \item Unlike in approaches based on factorisation theorems, in this paper, we have calculated the energy $E_J$, momentum $|\mathbf{P}_J|$ and mass $M_J$ distributions of jets produced in $e^+e^-$ annihilations with 2-- and 3--jet final states using \textbf{\textit{virtual leading partons}} attached to the previously derived TS-shaped FFs. The results show (Fig.~\ref{fig:jetDists}) that there is a hierarchy among the $\langle E_J\rangle$, $\langle |\mathbf{P}_J|\rangle$ and $\langle M_J\rangle$ of the jets depending on which generation the leading parton of the jet was produced at. Besides, we have found that the energy distribution of a jet, obtained using virtual leading partons and LL resummation in a 2-jet event, was well approximated by the distribution of an on-shell leading parton, whose jet pair contained only a single splitting (Fig.~\ref{fig:onOff}-\ref{fig:onOff2}). 
 
 \item Furthermore, when calculating hadronic distributions, we have found that the larger the jet mass, the more, and also more energetic hadrons stem from it. The angle between the momenta of hadrons and that of the jet is larger too (Fig.~\ref{fig:hadDists}). 
 
 \item We have developed an event generator to obtain the momenta of all hadrons produced in jets event-by-event, using the FF based on micro-canonical statistics and superimposed NBD hadron multiplicity fluctuations. We have derived that the mean multiplicity of hadrons in a jet depends on the jet mass as $\bar n \sim \ln^a(M_J)$ (see \REF{TS2}).
\end{itemize}

\section{Appendix}
\appendix
\section{Phasespace of $n$ massless particles}
\label{sec:micro}
\be
\Omega_n(P) \;=\; \prod_{i=1}^n \int \frac{d^{D-1}\mathbf{p}_i}{p_i^0}\, \delta^{D}\left(\sum_j p_j^\mu-P^\mu \right) \sim \int d^Ds\, e^{-is_\mu P^\mu} \varphi^n(s) \;,
\ee{A1}
where the Fourier-transform of the single-particle phasespace $\varphi(s)$ can be evaluated in the frame, in which, $s = (\sigma,\mathbf0)$ (with $\sigma^2 = s_0^2 - \mathbf{s}^2$), and $p = (p,\mathbf p)$:
\be
\varphi(\sigma) \;=\; \int \frac{d^{D-1}\mathbf{p}}{p^0}\, e^{is_\mu p^\mu} \sim \int dp\,p^{D-3} e^{i\sigma p} \sim \frac{1}{(i\sigma)^{D-2}} \;.
\ee{A2}
We may evaluate the inverse Fourier-transform in Eq.~(\ref{mic1}) in the frame, where $P = (M_0,\mathbf0)$, thus,
\be
\Omega_n(P) \;\sim\; \int d^{D-1}\mathbf{s} \int ds_0\, \frac{e^{-is_0 M_0} }{ [(s_0+|\mathbf s|)(s_0-|\mathbf s|)] ^{n(D-2)/2} }\;.
\ee{A3}
As we have poles at $s_0 = \pm|\mathbf s|$, we use Cauchy's formula $\oint \frac{dzf(z)}{(z-z_0)^n}\sim f^{(n-1)}(z_0)$, and arrive at terms of the form of
\ba
\Omega_n(P) &\;\sim\;& \sum_j A^{\pm}_j \int d^{D-1}\mathbf{s} \, \left(\frac{\partial}{\partial s_0}\right)^j  e^{-is_0 M_0}   \left(\frac{\partial}{\partial s_0}\right)^{n(D-2)/2-1-j} \left.\frac{1}{ (s_0\pm|\mathbf s|) ^{n(D-2)/2} } \right|_{s_0=\pm|\mathbf s|} \nl 
&& \sim \sum_j A^{\pm}_j\, M_0^j \int ds\, s^{D-1 - n(D-2)+j}  \, e^{-is M_0} \sim M_0^{n(D-2)-D}  \;.
\ea{A4}
The actual values of the constant factors $A^{\pm}_j$ multiplying the terms coming from the poles at $s_0 = \pm|\mathbf s|$, are of no importance from the point of view of the particle distributions.

\section{Calculation of the splitting function}
\label{sec:calcSF}
Via introducing the renormalized field and coupling $g=Z_g g_r$ and $\phi= Z_3^{1/2} \phi_r $, along with $Z_3 = 1 + \delta Z_3$ and $Z_g = 1 + \delta Z_g$, we arrive at the renormalised Lagrangian
\ba
\mathcal{L} &\;=\;& \frac{1}{2}(\partial_\mu\phi_r)^2 + (Z_3-1)\frac{1}{2}(\partial_\mu\phi_r)^2 + \frac{g_r}{3!}\phi_r^3  + (Z_g Z_3^{3/2}-1)\frac{g_r}{3!}\phi_r^3\nl
&=& \frac{1}{2}(\partial_\mu\phi_r)^2 + \delta Z_3\frac{1}{2}(\partial_\mu\phi_r)^2 + \frac{g_r}{3!}\phi_r^3  + (\delta Z_g + \frac{3}{2}\delta Z_3)\frac{g_r}{3!}\phi_r^3\;.
\ea{sf2}
As $\delta Z_3$ and $\delta Z_g$ are of $\mathcal{O}(g^2)$, terms proportional to them come as perturbative corrections. This way, a propagator of momentum $p$ is $i/(p^2+i\epsilon)$, a vertex is $-ig_r$, the counter terms are $ip^2\delta Z_3$ and $-ig_r(\delta Z_g+ \frac{3}{2}\delta Z_3)$, and a cut propagator is $2\pi\delta(p^2)$. Propagators and vertices on the right of the cuts are complex conjugates. 

We may write $A(z,P^2) = \delta(1-z)A_1(z,P^2) + A_2(z,P^2)$ in Eq.~(\ref{evol4}). When calculating $A_1(z,P^2)$, we use the identity\\ $\prod\limits_i \dfrac{1}{A_i^{\alpha_i}} = \dfrac{\Gamma(\alpha)}{\prod\limits_i \Gamma(\alpha_i)} \prod\limits_i\int\limits_0^1 \dfrac{d\xi_i\,\xi_i^{\alpha_i-1} \delta(1-\alpha)}{(\sum \xi_i A_i)^\alpha}$, with $\alpha = \sum\alpha_i$, thus,
\ba
A_1(z,P^2) &\;=\;& \int\limits_0^1 d\xi_1 \int\frac{d^Dq}{(2\pi)^D} \left\lbrace \int\limits_0^{1-\xi_1} d\xi_2      \frac{2i\,n_c}{\left[(1-\xi_1-\xi_2)q^2 + \xi_1(q-\hat k)^2 + \xi_2(P-q)^2\right]^3} \right.\;+\nl
&&\quad +\; \left.  \frac{i\,n_d}{m_0^2\left[(1-\xi_1)q^2 + \xi_1(q-\hat k)^2\right]^2}  + \frac{i\,n_e}{m_0^2 \left[(1-\xi_1)q^2 + \xi_1 (P-q-\hat k)^2\right]^2}  \right\rbrace\nl
&+& \frac{\bar n_c}{g^2} \left(\delta Z_g + \frac{3}{2}\delta Z_3\right) - \frac{\bar n_d + \bar n_e}{g^2}\delta Z_3  \;.
\ea{1p19}
Substituting $\tilde q=q - \xi_1k-\xi_2P$ and $L_1 = -P^2\xi_2(1-\xi_2)$ in the first term, $\tilde q = q-\xi_1k$ in the second term and $\tilde q = q-\xi_1(P-k)$ in the third term, we get
\ba
g^2 A_1(z,P^2) &\;=\;& i\,g^2 \int\limits_0^1 d\xi_1 \int\frac{d^D\tilde q}{(2\pi)^D} \left\lbrace  \int\limits_0^{1-\xi_1} d\xi_2    \frac{2n_c}{(\tilde q^2 - L_1)^3} \;+\;  \frac{n_d + n_e}{m_0^2\, \tilde q^4}  \right\rbrace +\nl
&&+\; \bar n_c \left(\delta Z_g + \frac{3}{2}\delta Z_3\right) - (\bar n_d + \bar n_e)\delta Z_3 \;.
\ea{1p20}
Substituting $\tilde q = (iq_E,\mathbf q_E)$ (Wick-rotation) gives 
\ba
g^2A_1(z,P^2) &\;=\;& g^2\frac{\kappa_D}{(2\pi)^D} \int\limits_0^1 d\xi_1 \int\limits_0^\infty dq_E\,q_E^{D-1} \left\lbrace  \int\limits_0^{1-\xi_1} d\xi_2    \frac{2n_c}{( q^2_E + L_1)^3} \;-\;  \frac{n_d + n_e}{m_0^2\, \tilde q_E^4}e^{-\epsilon q_E/m_0}  \right\rbrace  +\nl
&&+\; \bar n_c \left(\delta Z_g + \frac{3}{2}\delta Z_3\right) - (\bar n_d + \bar n_e)\delta Z_3 \;.
\ea{1p21}
Using $\int\limits_0^\infty\dfrac{dx\,x^{a-1}}{(x+1)^{a+b}} = \Gamma(a)\Gamma(b)/\Gamma(a+b)$, $\Gamma(\epsilon) = 1/\epsilon-\gamma_E$ and setting $D=6-2\epsilon$ along with $g\rightarrow g\mu^\epsilon$, we obtain
\ba
g^2A_1(z,P^2) &\;=\;& \frac{g^2 n_c}{2 (4\pi)^3}   \left(\frac{1}{\epsilon} - \ln\frac{-P^2}{\mu^2}  -\gamma_E + \ln4\pi  - 2\int\limits_0^1 d\xi_1 \int\limits_0^{1-\xi_1} d\xi_2 \ln \xi_2(1-\xi_2) \right) \;-\nl 
&& - \frac{g^2 }{2 (4\pi)^3} (n_d + n_e)\frac{1}{\epsilon^2}    +\; \bar n_c \left(\delta Z_g + \frac{3}{2}\delta Z_3\right) - (\bar n_d + \bar n_e)\delta Z_3 \;.
\ea{1p22}
If we remove the divergences along with the $P^2$ and $\mu$ independent constants by the counter terms, we are left with
\be
A_1(z,P^2) \;=\; -\frac{n_c }{2 (4\pi)^3} \ln\frac{P^2}{\mu^2}
\ee{1p23}

For the calculation of the second line of $A(z,P^2)$ in Eq.~(\ref{evol4}), we parametrize momenta as $P=(M,0,\mathbf0)$, $k=(Mz/2,Mz/2,\mathbf0)$ and $q=\alpha P + \beta k + q_T = (\alpha M + \beta Mz/2,\beta Mz/2,\mathbf q_T)$. This way, the integration measure becomes $d^Dq = (M^2z/2)d\alpha\, d\beta\, d^{D-2}\mathbf q_T$. Besides, $2Pq = M^2(2\alpha + \beta z)$, $2 k q = M^2\alpha z$ and $q^2 = M^2\alpha(\alpha+\beta z)-\mathbf q_T^2$. Furthermore, due to the $\delta\left[(q-k)^2\right]$ term, $q^2 = 2k q = M^2\alpha z$. This way, the second line of Eq.~(\ref{evol4}) becomes
\ba
g^2A_2(z,P^2) &\;=\;& g^2\frac{z \kappa_{D-2}}{4(2\pi)^{D+1}} \int d\alpha \int d\beta \int dq_T^2\,(q_T^2)^{D/2-2} \;\times\nl
&&\times \; (2\pi)\delta\left[M^2\alpha(\alpha + \beta z - z)-q_T^2\right] (2\pi)\delta\left[M^2(1+\alpha z - 2\alpha -\beta z)\right] \;\times\nl
&&\times\; \left\lbrace \frac{n_f}{\alpha^2 z^2 } + \frac{n_g}{(1-z)^2} + \frac{n_h}{ \alpha z (1-z)} + \frac{n_i}{\alpha z  (1-2\alpha+z-\beta z)} \right\rbrace \;.\nl
\ea{1p30}
Using the integral for $\beta$ to eliminate the second $\delta$ function gives
\ba
g^2 A_2(z,P^2) &\;=\;& \frac{g^2\kappa_{D-2}}{4(2\pi)^{D-1} M^2} \int d\alpha \int dq_T^2\,(q_T^2)^{D/2-2}  \delta\left[M^2\alpha(1-\alpha)(1 - z)-q_T^2\right] \;\times\nl
&&\times\; \left\lbrace \frac{n_f}{\alpha^2 z^2 } + \frac{n_g}{(1-z)^2} + \frac{n_h}{ \alpha z (1-z)} + \frac{n_i}{\alpha(1-\alpha)z^2} \right\rbrace \nl
&\;=\;& \frac{g^2\kappa_{D-2} M^{D-6} (1-z)^{D/2-2}}{4(2\pi)^{D-1}} \int d\alpha \,[\alpha(1-\alpha)]^{D/2-2}   \;\times\nl
&&\times\; \left\lbrace \frac{n_f}{\alpha^2 z^2 } + \frac{n_g}{(1-z)^2} + \frac{n_h}{ \alpha z (1-z)} + \frac{n_i}{\alpha(1-\alpha)z^2} \right\rbrace \nl
\ea{1p31}

Using the identity $\int\limits_0^1 d\alpha \alpha^{a-1}(1-\alpha)^{b-1} = \Gamma(a)\Gamma(b)/\Gamma(a+b)$ and $D=6-2\epsilon$ dimensions, where the coupling acquires dimension $g\rightarrow g\mu^\epsilon$, the solid angle is $\kappa_D = 2\pi^{D/2}/\Gamma(D/2)$ and $\Gamma(-\epsilon)=-1/\epsilon-\gamma_E$, we obtain

\ba
g^2 A_2(z,P^2) &\;=\;& \frac{g^2 (1-z)}{(4\pi)^3} \left(\frac{4\pi\mu}{M^2(1-z)} \right)^\epsilon
 \left\lbrace \frac{n_f}{z^2}\left(-\frac{1}{\epsilon}-\gamma_E \right) + \frac{n_g}{6(1-z)^2} + \frac{n_h}{2z (1-z)} + \frac{n_i}{z^2} \right\rbrace \nl
&\;=\;& \frac{g^2}{(4\pi)^3} 
 \left\lbrace -\left[\frac{1}{\epsilon}+\gamma_E + \ln\left(\frac{4\pi\mu}{M^2(1-z)} \right) \right]\frac{n_f(1-z)}{z^2 }\; + \right. \nl
&& \qquad\qquad + \left. \frac{n_g}{6(1-z)} + \frac{n_h}{2z} + \frac{n_i(1-z)}{z^2} \right\rbrace \;.
\ea{1p32}
We made use of $\Gamma(-1+\epsilon)=-1/\epsilon+\gamma_E-1$, $\Gamma(-\epsilon)=-1/\epsilon-\gamma_E$. Note that the $1/\epsilon$ term being the collinear divergence, cannot be eliminated via renormalisation, however, it drops out of the splitting function, which is 
\be
\Pi(z) \;=\; \frac{\partial}{\partial\ln P^2}A(z,P^2) \;=\; \frac{n_f}{(4\pi)^3}\frac{1-z}{z^2}  - \frac{n_c}{2 (4\pi)^3} \delta(1-z) \;.
\ee{1p33}

%\begin{figure}
%\resizebox{0.75\textwidth}{!}{%
%  \includegraphics{leer.eps}
%\vspace{5cm}       % Give the correct figure height in cm
%\end{figure}
%\begin{figure*}
%\vspace*{5cm}       % Give the correct figure height in cm
%\end{figure*}
%\begin{table}
%\begin{tabular}{lll}
%\hline\noalign{\smallskip}
%\end{tabular}
% Or use
%\vspace*{5cm}  % with the correct table height
%\end{table}
%
% BibTeX users please use
% \bibliographystyle{}
% \bibliography{}

\twocolumn[]

\end{document}